\magnification=\magstep1


\def\ou{Ornstein-Uhlenbeck process}
\def\pd{probability distribution}

\def\ito{It\^o}

\def\real{{\pmb{R}}}

\def\ep{\epsilon}

\def\damtp{\baselineskip=10pt
\eightsl Department of Applied Mathematics and Theoretical Physics,\break%
\eightsl University of Cambridge, Cambridge CB3 9EW, UK\par}
\def\ont{\baselineskip=10pt
\eightsl Optique nonlin\'eaire th\'eorique, \break
Universit\'e libre de Bruxelles, CP231 \break
Bruxelles 1050 BELGIUM}
\def\and{\ifnyasrefstyle \&\ \else and \fi}%
\def\regime{r\'egime}%
%
%

\def\mle{\mu |\ln \epsilon |}
\def\Or#1{{\cal O}(#1)}
\def\mean#1{{\big<#1\big>}}

\def\ha{{1 \over 2}}



\font\sixrm=cmr6           \font\sevenrm=cmr7
\font\sixi=cmmi6           \font\seveni=cmmi7
\font\sixsy=cmsy6          \font\sevensy=cmsy7

\font\eightrm=cmr8         \font\tenrm=cmr10
\font\eighti=cmmi8         
\font\eightsy=cmsy8        
\font\eightit=cmti8        \font\tenit=cmti10
\font\eightsl=cmsl8        \font\tensl=cmsl10
\font\eightbf=cmbx8        \font\tenbf=cmbx10
\font\eighttt=cmtt8        

\font\twelverm=cmr12

\font\twelveit=cmti12
\font\twelvesl=cmsl12
\font\twelvebf=cmbx12

\font\fourteenrm=cmr12 at 14pt

\font\fourteenit=cmti12 at 14pt
\font\fourteensl=cmsl12 at 14pt
\font\fourteenbf=cmbx12 at 14pt




\def\pmb#1{\setbox0=\hbox{#1}%
 \kern-.025em\copy0\kern-\wd0%
 \kern.05em\copy0\kern-\wd0%
 \kern-.025em\raise.0433em\box0}
\def\Bigtype{%
\let\rm=\fourteenrm \let\bf=\fourteenbf%
\let\it=\fourteenit \let\sl=\fourteensl%
\baselineskip=1.44\baselineskip \rm}%
\def\bigtype{%
\let\rm=\twelverm \let\bf=\twelvebf%
\let\it=\twelveit \let\sl=\twelvesl%
\baselineskip=1.2\baselineskip \rm}%
\def\medtype{%
\let\rm=\tenrm \let\bf=\tenbf%
\let\it=\tenit \let\sl=\tensl%
\baselineskip=12pt \rm}%
\def\smalltype{%
\let\rm=\eightrm \let\bf=\eightbf%
\let\it=\eightit \let\sl=\eightsl%
\let\tt=\eighttt
\def\cal{\fam2 }%
 \textfont0=\eightrm   \scriptfont0=\sevenrm \scriptscriptfont0=\sixrm
 \textfont1=\eighti    \scriptfont1=\seveni  \scriptscriptfont1=\sixi
 \textfont2=\eightsy   \scriptfont2=\sevensy \scriptscriptfont2=\sixsy
 \textfont3=\tenex     \scriptfont3=\tenex   \scriptscriptfont3=\tenex
 \textfont\itfam=\eightit
 \textfont\slfam=\eightsl
 \textfont\bffam=\eightbf
 \textfont\ttfam=\eighttt
\baselineskip=10pt
\parskip=1pt plus1pt minus 1pt%
\rm}%
\voffset=0truein%
\vsize=9.3truein%
\hoffset=0.1truein%
\hsize=6.0truein%
\parindent=0.375truein%
\widowpenalty=300

\def\spaceandahalf{%
\baselineskip=15pt
\parskip=4pt plus4pt minus2pt%
\everydisplay={\openup 1.5\jot}}%

\def\singlespace{%
\baselineskip=12pt
\parskip=1pt plus1pt minus 1pt%
\everydisplay={\openup 0\jot}}%

\def\closeup{\singlespace}%

\newdimen\pagewidth
\newdimen\pageheight    
\pagewidth=6.0truein
\vsize=9.3truein%
\pageheight=9.3truein
\newif\ifwindowcover%
\windowcoverfalse%
\outer\def\windowcover{\windowcovertrue}%
\newif\ifcam%
\camfalse%
\outer\def\damtpaddress{\camtrue}%
\newif\ifnofigs%
\nofigsfalse%
\newif\ifchapstart
\chapstartfalse
\def\chapstartpage{\global\chapstarttrue}
\newif\ifinappendices%
\inappendicesfalse%
\newif\iffiguresatend%
\newif\ifnumbereqnsseq%
\def\numbereqnsseq{ \numbereqnsseqtrue}%
\newif\ifnumberfigsseq%
\def\numberfigsseq{ \numberfigsseqtrue}%
\newif\ifnumbertablesseq%
\def\numbertablesseq{ \numbertablesseqtrue}%
\newif\ifnumberrefsintext%
\def\numberrefsintext{ \numberrefsintexttrue}%
\newif\ifinrefs%
\inrefsfalse%
\newif\ifpaperstyle%
\outer\def\paperstyle{\paperstyletrue
     \refauthorinitialssurname%
     \numbereqnsseq \numberfigsseq \numbertablesseq%
     \numberrefsintext}%
\newif\ifsendstyle%
\newif\ifnyasrefstyle%

\newif\ifsinglesided%
\newif\ifdated%

\newif\ifsastyle%
\outer\def\sastyle{\sastyletrue\numberrefsintext%
\ifsinglesided\headline={\rightheadline}
\else\headline={\ifodd\pageno\rightheadline
                        \else\leftheadline\fi}\fi%
\footline={\ifdated\hfil\eightrm\today\else{}\fi}}
\newif\ifheadset%
\headsetfalse%

\countdef\sectno=1%
\countdef\subsectno=2%
\countdef\subsubsectno=3%
\newcount \equano%
\newcount \figno%
\newcount \tableno%
\newcount \referencecount%
\newbox\TitlePage%
\newbox\DedicationPage%
\newbox\PrefacePage%
\newbox\ConclusionPage%
\newbox\AcknowledgementsPage%
\newbox\SummaryPage%
\newbox\TableOfContents%
\newbox\ListOfFigures%
\newbox\Nomenclature%
\newbox\ListOfTables%
\newbox\FiguresPage%
\newbox\figurebox%
\def\thesectno{%
\ifinappendices%
  \ifcase\sectno\or%
    A\or B\or C\or D\or E\or F\or G\or H\or I\or J\or K\or L\or M%
    N\or O\or P\or Q\or R\or S\or T\or U\or V\or W\or X\or Y\or Z\fi%
\else%
  \the\sectno%
\fi}%
\def \lastsubsectno{\thesectno .\the\subsectno}%
\def \lastsubsubsectno{\thesectno .\the\subsectno .\the\subsubsectno}%
\def \lasteqno{\ifnumbereqnsseq (\the\equano)%
\else (\thesectno.\the\equano)\fi}%
\def \preveqno{\advance \equano by -1 \lasteqno}%
\def \lastfigno{\ifnumberfigsseq \hbox{Figure}~\hbox{\the\figno}%
\else \hbox{Figure}~\hbox{\thesectno -\the\figno}\fi}
\def \lasttableno{\ifnumbertablesseq \hbox{Table}~\hbox{\the\tableno}%
\else \hbox{Table}~\hbox{\thesectno .\the\tableno}\fi}%
\def \nextsubsectno{{\advance \subsectno by 1 \lastsubsectno}}
\def \nextsubsubsectno{{\advance \subsubsectno by 1 \lastsubsubsectno}}%
\def \nexteqno{{\advance \equano by 1 \lasteqno}}%
\def \nextfigno{{\advance \figno by 1 \lastfigno}}%
\def \nextnextfigno{{\advance \figno by 2 \lastfigno}}%
\def \nexttableno{{\advance \tableno by 1 \lasttableno}}%
\def\minimalsectno{\ifnum\subsubsectno>0 \lastsubsubsectno%
\else\ifnum\subsectno>0 \lastsubsectno%
\else \thesectno \fi\fi}%
\def\beginsect#1\par{\begingroup
\ifsastyle\thesisheading{#1}
        \chapstartpage
    \xdef\rhead{\ #1}
\else%
    \global \advance \sectno by 1%
    \mainheading{#1}%
\fi
\global \subsectno=0 \global \subsubsectno=0%
\ifnumbereqnsseq \else \global \equano=0 \fi%
\ifnumberfigsseq \else \global \figno=0 \fi%
\ifnumbertablesseq \else \global \tableno=0 \fi%
\message{\thesectno . #1}%
\nobreak\endgroup%
\ifsastyle\headsetfalse\else\fi}%

\def\saheading#1{\vfill\supereject
        \global \advance \sectno by 1%
        \chapstartpage
        \vskip 0pt plus.3\vsize \penalty-100 \vskip 0pt plus-.3\vsize%
        \ifinrefs\noindent{\bigtype \bf #1}
        \else\ifinappendices\noindent{\bigtype \bf #1}
        \else\noindent{\bigtype \the\sectno .\ \bf #1}\fi\fi
        \par\nobreak\noindent}

\def\thesisheading#1{\vfill\supereject
        \global \advance \sectno by 1%
        \vskip 0pt plus.3\vsize \penalty-100 \vskip 0pt plus-.3\vsize%
        \ifinrefs\noindent{\bigtype \bf #1}
        \else\ifinappendices\noindent{\bigtype \bf Appendix \thesectno .\ #1}
        \else\line{}\vskip 0.5truein
        \centerline{\twelverm CHAPTER \the\sectno}
                \vskip 0.5truein
        \centerline{\fourteenbf #1}
        \vskip 0.5truein
        \fi\fi
        \par\nobreak\noindent}

\def\mainheading#1{\begingroup\nohyphens\hbadness=3000%
        \ifinrefs%
          \noindent {\bf #1}\nobreak%
        \else%
           \ifinappendices%
              \noindent {\bf #1}\par\nobreak%
           \else
              \noindent {\bf \thesectno . #1}\par\nobreak%
           \fi%
        \fi%
    \noindent\endgroup}%

\def\subheading#1{\begingroup\nohyphens\hbadness=3000%
\bigskip
    \noindent {\sl\bf\lastsubsectno . #1}\par\nobreak\noindent%
\endgroup}%

\def\subsubheading#1{\begingroup\nohyphens\hbadness=3000%
  \vskip 0pt plus.1\vsize \penalty-100 \vskip 0pt plus-.1\vsize%
\medskip\vskip\parskip%
  \noindent {\sl\lastsubsubsectno . #1}\par\nobreak\smallskip\noindent%
\endgroup}%

\outer \def \beginappendix#1\par{%
\inrefsfalse%
\ifinappendices\else\inappendicestrue\global\sectno=0\fi%
\begingroup%
\ifsastyle\saheading{#1}%
\xdef\rhead{#1}
\else%
    \global \advance \sectno by 1%
    \mainheading{#1}%
    \ifnumbereqnsseq \else \global \equano=0 \fi%
    \ifnumberfigsseq \else \global \figno=0 \fi%
    \ifnumbertablesseq \else \global \tableno=0 \fi%
    \global \subsectno=0 \global \subsubsectno=0%
\fi%
\message{\thesectno . #1}%
\endgroup%
\ifsastyle\headsetfalse\else\fi%
}

\outer \def \beginsubsect#1\par{
\global \advance \subsectno by 1%
\global \subsubsectno=0%
\accretetoTOCtwo{\lastsubsectno.}{#1}%
\subheading{#1}}

\outer \def \beginsubsubsect#1\par{\global \advance \subsubsectno by 1%
\accretetoTOCthree{\lastsubsubsectno.}{#1}\subsubheading{#1}}%
\def \myeqno{\global \advance \equano by 1 {\rm \lasteqno}}%
\def \myeqlabel#1{\expandafter\xdef\csname #1\endcsname{\lasteqno}}

\def \mytextlabel#1{\expandafter\xdef\csname #1\endcsname{\minimalsectno}}
\def \myChapterlabel#1#2{%
 \expandafter\xdef\csname #1\endcsname{Chapter~#2}}
\def \myAppendixlabel#1#2{%
 \expandafter\xdef\csname #1\endcsname{Appendix~\hbox{#2}}}
\def \myfigurelabel#1{\expandafter\xdef\csname #1\endcsname{\lastfigno}}
\def \mytablelabel#1{\expandafter\xdef\csname #1\endcsname{\lasttableno}}
\input epsf

\def\figure #1 #2 #3\par{\global \advance \figno by 1%
\iffiguresatend%
   \global\setbox\ListOfFigures=\vbox{\unvbox\ListOfFigures%
   \bigskip\caption{\hbox{\bf\lastfigno. }{#3}}
   \par\filbreak}%
   \ifnofigs%
   \else
      \global\setbox\FiguresPage=\vbox{\unvbox\FiguresPage%
        \epsfysize=#1
        \centerline{\epsffile{#2}}
       \vskip 0.5truein\relax%
       \hbox{\bf \lastfigno}
        {#3}
        \vfill\eject}
   \fi%
\else%
\ifsastyle
   \global\setbox\ListOfFigures=\vbox{\unvbox\ListOfFigures%
   \bigskip\caption{\hbox%
{\bf\lastfigno. }
{page \folio},
{size in text #1},
{file #2}\filbreak\noindent{#3}}
   \par\filbreak}%
\fi
  \setbox\figurebox=\vbox{%
  \ifnofigs
  \else
        \epsfysize=#1   
        \centerline{\epsffile{#2}}
  \fi
    \caption{{\bf\lastfigno. }{\eightsl #3}}}
  \topinsert\unvbox\figurebox\endinsert%
\fi}%

\def\twofigures #1 #2 #3 #4\par{\global \advance \figno by 1%
\iffiguresatend 
   \begingroup%
   \global\setbox\ListOfFigures=\vbox{\unvbox\ListOfFigures%
   \bigskip\caption{\hbox{\bf\lastfigno. }{#4}}
   \par\filbreak}%
   \ifnofigs
   \else
      \global\setbox\FiguresPage=\vbox{\unvbox\FiguresPage%
        \epsfysize=#1
        \centerline{\epsffile{#2}\hfil\epsffile{#3}}
       \vskip 0.5truein\relax%
       \hbox{\bf \lastfigno}
        \vfill\eject}
   \fi
   \endgroup%
\else%
  \setbox\figurebox=\vbox{%
  \ifnofigs
  \else
        \centerline{
        \epsfysize=#1   
        (a)\epsffile{#2}
        \ \ 
        \epsfysize=#1   
        (b)\epsffile{#3}}
  \fi
    \caption{{\bf\lastfigno. }{\eightsl #4}}}
  \topinsert\unvbox\figurebox\endinsert%
\fi}%

\def\fourfigures #1 #2 #3 #4 #5 #6\par{\global \advance \figno by 1%
\iffiguresatend 
   \begingroup%
   \global\setbox\ListOfFigures=\vbox{\unvbox\ListOfFigures%
   \bigskip\caption{\hbox{\bf\lastfigno. }{#6}}
   \par\filbreak}%
   \ifnofigs
   \else
      \global\setbox\FiguresPage=\vbox{\unvbox\FiguresPage%
        \centerline{
        \epsfysize=#1   
        (a)\epsffile{#2}
        \ \ 
        \epsfysize=#1   
        (b)\epsffile{#3}}
        \line{ }
        \centerline{
        \epsfysize=#1   
        (c)\epsffile{#4}
        \ \ 
        \epsfysize=#1   
        (d)\epsffile{#5}}
        \hbox{\bf \lastfigno}
         {#6}
        \vfill\eject}
   \fi
   \endgroup%
\else%
  \setbox\figurebox=\vbox{%
  \ifnofigs
  \else
        \centerline{
        \epsfysize=#1   
        (a)\epsffile{#2}
        \ \ 
        \epsfysize=#1   
        (b)\epsffile{#3}}
        \line{ }
        \centerline{
        \epsfysize=#1   
        (c)\epsffile{#4}
        \ \ 
        \epsfysize=#1   
        (d)\epsffile{#5}}
  \fi
    \caption{{\bf\lastfigno. }{\eightsl #6}}}
  \topinsert\unvbox\figurebox\endinsert%
\fi}%

\def\eightfigures #1 #2 #3 #4 #5 #6 #7 #8 #9%
\par{\global \advance \figno by 1%
\iffiguresatend 
   \begingroup%
   \global\setbox\ListOfFigures=\vbox{\unvbox\ListOfFigures%
   \bigskip\caption{\hbox{\bf\lastfigno. }{#6}}
   \par\filbreak}%
   \ifnofigs
   \else
      \global\setbox\FiguresPage=\vbox{\unvbox\FiguresPage%
        \epsfysize=#1
        \centerline{\epsffile{#2}\hfil\epsffile{#3}}
       \vskip 0.5truein\relax%
       \hbox{\bf \lastfigno}
        \vfill\eject}
   \fi
   \endgroup%
\else%
  \setbox\figurebox=\vbox{%
  \ifnofigs
  \else
        \centerline{
        \epsfysize=1.8truein
        (a)\epsffile{#1}
        \epsfysize=1.8truein
        (b)\epsffile{#2}
        \epsfysize=1.8truein
        (c)\epsffile{#3}}
        \line{ }
        \centerline{
        \epsfysize=1.8truein
        (d)\epsffile{#4}
        \epsfysize=1.8truein
        (e)\epsffile{#5}
        \epsfysize=1.8truein
        (f)\epsffile{#6}}
        \line{ }
        \centerline{
        \epsfysize=1.8truein
        (g)\epsffile{#7}
        \epsfysize=1.8truein
        (h)\epsffile{#8}
        \epsfysize=1.8truein
        \ \ \ \epsffile{key.eps}}
  \fi
    \caption{{\bf\lastfigno. }{\eightsl #9}}
}
  \topinsert\unvbox\figurebox\endinsert%
\fi}%

\def\caption#1{%
\smallskip%
\begingroup
\nohyphens%
\advance \leftskip by \parindent%
\advance \rightskip by \parindent
\spaceskip=.3333em plus .3333em \xspaceskip=.5em plus .5em%
\par \noindent{%
\iffiguresatend{#1}\else\smalltype{#1}\par\fi\par} \par\endgroup}%

\outer \def \beginrefs{\inrefstrue\inappendicesfalse}%

\def\showrefs{\ifsastyle\accretetoTOCone{}{References}\headsetfalse\else\fi%
\setbox\myrefs=\vbox{\unvbox\myrefs\vfil}%
\ifsendstyle\vfill\eject\else\smallskip\goodbreak\fi
    \unvbox\myrefs}

\def\addtorefs#1#2{%
\global\advance\referencecount by 1%
\expandafter\xdef\csname#2\endcsname{\the\referencecount}%
\global\setbox\myrefs=\vbox{\unvbox\myrefs%
{\ifsendstyle\else\closeup\fi
\ifnyasrefstyle
\item{\csname#2\endcsname.}\frenchspacing{\csname#1\endcsname}
\else
\item{[\csname#2\endcsname]}\frenchspacing{\csname#1\endcsname}
\fi
\hfil\par\ifsendstyle\medskip\fi}}}

\def\addref#1#2{\ifundefined{#2}\addtorefs{#1}{#2}\fi
\ifnyasrefstyle$^{\csname#2\endcsname}$\else~[\csname#2\endcsname]\fi}%


\def\addtworefs#1#2#3#4{%
{\ifundefined{#2}\addtorefs{#1}{#2}\fi}%
{\ifundefined{#4}\addtorefs{#3}{#4}\fi}%
\ifnyasrefstyle
$^{\csname#2\endcsname,~\csname#4\endcsname}$%
\else~[\csname#2\endcsname,~\csname#4\endcsname]\fi}%

\def\addthreerefs#1#2#3#4#5#6{%
{\ifundefined{#2}\addtorefs{#1}{#2}\fi}%
{\ifundefined{#4}\addtorefs{#3}{#4}\fi}%
{\ifundefined{#6}\addtorefs{#5}{#6}\fi}%
\ifnyasrefstyle
$^{\csname#2\endcsname,\csname#4\endcsname,\csname#6\endcsname}$%
\else~[\csname#2\endcsname,\csname#4\endcsname,\csname#6\endcsname]\fi}%

\def\addfourrefs#1#2#3#4#5#6#7#8{
{\ifundefined{#2}\addtorefs{#1}{#2}\fi}%
{\ifundefined{#4}\addtorefs{#3}{#4}\fi}%
{\ifundefined{#6}\addtorefs{#5}{#6}\fi}%
{\ifundefined{#8}\addtorefs{#7}{#8}\fi}%
\ifnyasrefstyle
$^{\csname#2\endcsname-\csname#8\endcsname}$%
\else~[\csname#2\endcsname-\csname#8\endcsname]\fi}%


\def\genericref#1#2{\begingroup\closeup%
\expandafter\xdef\csname#1\endcsname{#2.}%
\endgroup}%
\def\refyear#1{\ifnyasrefstyle #1. \else (#1)\fi}%
\def\reftitle#1{\lq\lq#1''}%
\def\refauthorinitialssurname{%
\def\author##1##2{##1~##2}
\def\authorn##1##2{\ifnyasrefstyle\uppercase{##2,~##1}\else{##1~##2}\fi}}%
\refauthorinitialssurname%
\outer\def\refarticle#1#2#3#4#5#6#7{%
\genericref{#1}{%
\ifnyasrefstyle
    {#2.\ \refyear{#3} \reftitle{#4}. {#5} {\bf #6}: {#7}}%
\else%
  \ifpaperstyle%
    {#2\ \reftitle{#4.} {\sl #5} {\bf #6} #7\ \refyear{#3}}%
  \else
    {#2\ \reftitle{#4.} {\sl #5} {\bf #6} #7\ \refyear{#3}}%
  \fi%
\fi}}%

\outer\def\refarticletobe#1#2#3#4#5{%
\genericref{#1}{#2, \reftitle{#4} {#5} \refyear{#3}}}

\outer\def\refbook#1#2#3#4#5#6{%
\genericref{#1}%
{\ifnyasrefstyle%
#2.\ #3. #4. #5, #6%
\else%
\ifpaperstyle%
{#2,\ {\sl #4}\ (#5, \ #6, \ #3)}%
  \else%
    {#2\ {\sl #4}.  #5: #6 \ \refyear{#3}}%
\fi\fi}}%

\outer\def\refartbook#1#2#3#4#5#6#7#8#9{%
\genericref{#1}%
{\ifpaperstyle%
    {#2 \refyear{#3} \reftitle{#4,} in {\sl #5}, #6 (Ed.). #7: #8}%
\else%
    {#2 \reftitle{#4} #9 in {\sl #5}, Ed. #6. #7: #8 \refyear{#3}}%
\fi}}%

\outer\def\refphd#1#2#3#4#5{%
\genericref{#1}%
{\ifpaperstyle%
  {#2, {\sl #4} (Ph.D. thesis, #5, #3)}%
\else%
    {#2 \refyear{#3} \reftitle{#4.} {#5, Ph.D. thesis}}%
\fi}}

\outer\def\refnote#1{%
\genericref{#1}}
\def\theauthor{Grant Lythe}

\def\title#1\par{\message{#1}%
\xdef\lhead{#1}%
\xdef\rhead{}%
\begingroup
  \ifwindowcover%
    \dimen1=2.5truein \advance\dimen1 by -\voffset%
    \global\setbox\TitlePage=\vbox{\vbox to \dimen1{%
      \dimen2=1.75truein \advance\dimen2 by -\voffset%
      \smallskip \relax%
      \dimen3=1.5truein \advance\dimen3 by -\hoffset%
      \leftskip=\dimen3 plus3em\relax%
      \dimen4=\hsize \advance\dimen4 by \hoffset \advance\dimen4 by -5truein%
      \rightskip=\dimen4 plus3em\relax%
      \parfillskip=0pt \parindent=0pt \spaceskip=0.3333em
        \xspaceskip=0.5em%
        \line{}\bigskip\line{}\bigskip
      \centerline{\bigtype\bf #1}\smallskip
          {\centerline{\theauthor}}
        }}
  \fi%
\endgroup}%

\def\signaturepaper{
\global\setbox\TitlePage=\vbox{%
\unvbox\TitlePage%
\noindent\narrower\raggedcenter{\bigtype%
\ifwindowcover \else \theauthor\par\fi%
\ifcam\damtp\else\ont\fi
\bigskip}}}%

\def\signaturethesis{%
\parskip=20pt plus 10pt%
\global\setbox\TitlePage=\vbox{%
\line{}\bigskip
\vfill%
\noindent\narrower\raggedcenter{%
\bigtype%
{\bf Stochastic slow-fast dynamics}\par%
\bigtype%
Grant David Lythe\par%
\medtype
Trinity College, Cambridge\par%
\vfill%
Dissertation submitted\break%
for the degree of\break%
Doctor of Philosophy\break%
at the\break%
University of Cambridge\break%
\vfill%
October 1994}}}%

\def\signature{\begingroup%
\ifpaperstyle \signaturepaper \fi%
\ifsastyle \signaturethesis \fi%
\endgroup}%

\long\def\addtotitlepage#1{
\global\setbox\TitlePage=\vbox{%
\unvbox\TitlePage%
\bigskip\noindent#1\par}}%

\def\abstract#1\par{\begingroup%
\addtotitlepage{{\centerline{\bf Abstract}}\par\nobreak\medskip\noindent#1}%
\endgroup}%

\def\submittedto#1{\begingroup\closeup%
\addtotitlepage{\smalltype
\ifpaperstyle  {Submitted to  \hfill printed on \ \today}\fi%
}\endgroup}%

\def\publishedin#1{\begingroup\closeup%
\addtotitlepage{\bigtype
\ifpaperstyle  {#1}\fi}\endgroup}%

\long\def\summary#1{%
\global\setbox\SummaryPage=\vbox{%
  \unvbox\SummaryPage \noindent #1\par}%
  \begingroup\pageno=-4\endgroup}

\outer \def \beginack{
\noindent{\bf{Acknowledgement}}\break}

\def\Showbox#1{\begingroup%
\loop%
\ifvbox#1%
    \dimen255=\vsize%
    \advance \dimen255 by -\topskip%
    \setbox0=\vsplit#1 to \dimen255%
    \unvbox0%
    \null \vfill \penalty -10000%
\repeat%
\endgroup}%

\def\ShowTP{\begingroup \nopagenumbers\headline={}%
\ifvbox\TitlePage%
    \global \sectno=0 \global \subsectno=0 \global \subsubsectno=0%
    \setbox\TitlePage=\vbox{\unvbox\TitlePage\vfil}%
    \Showbox\TitlePage%
\fi \endgroup}%

\def\PreambleHeader#1{%
\begingroup\raggedcenter\line{}\vskip 0.3truein 
\ifsastyle\bigtype\else\Bigtype\fi\bf \noindent%
#1\par \vskip 0.2truein\endgroup\noindent}%

\def\ShowAck{%
\ifvbox\AcknowledgementsPage%
    \global \sectno=0 \global \subsectno=0 \global \subsubsectno=0%
    \setbox\AcknowledgementsPage=\vbox{%
\PreambleHeader{Acknowledgements}\unvbox\AcknowledgementsPage \vfil}%
    \Showbox\AcknowledgementsPage%
\fi}%

\def\ShowSummary{%
\ifvbox\SummaryPage%
    \global \sectno=0 \global \subsectno=0 \global \subsubsectno=0%
    \setbox\SummaryPage=\vbox{%
\PreambleHeader{Summary}\unvbox\SummaryPage \vfil}%
    \Showbox\SummaryPage%
\fi}%

\def\Trailers{%
\pageno=-1%
\footline={}
\ifsastyle\headline={}%
 \ShowTOC \fi
\iffiguresatend \begingroup \nopagenumbers \ShowLOF \ShowFP \endgroup\fi%
}%
\def\themonth{\ifcase\month\or%
  January\or February\or March\or April\or%
  May\or June\or July\or August\or%
  September\or October\or November\or December\fi}%
\def\today{\number\day%
\space%
\themonth%
\space%
\number\year}%
%
\outer\def\raggedright{\rightskip=0pt plus 4em%
  \spaceskip=.3333em \xspaceskip=.5em%
  \pretolerance=200 \tolerance=400}%
\def\leaderfill{\leaders\hbox to 1em{\hss.\hss}\hfill}%
\def\dddot#1{\begingroup\vbox{%
\moveleft 0.15em\hbox{...}%
\nointerlineskip\vskip0.4ex%
\hbox{$\displaystyle{#1}$}}\endgroup}%
\def\ifundefined#1{\expandafter\ifx\csname#1\endcsname\relax}%
\hyphenation{co-dimen-sion sol-ution sol-utions sun-spot sun-spots
pro-pen-sity}
\overfullrule=0pt%
\def\nohyphens{\pretolerance=9999 \tolerance=9999%
\hyphenpenalty=9999 \exhyphenpenalty=1000 }%
\def\raggedcenter{\leftskip=0pt plus6em \rightskip=\leftskip%
\parfillskip=0pt \parindent=0pt \spaceskip=0.3333em \xspaceskip=0.5em%
\nohyphens \hbadness=3000}%
%
%
%
%
%
\outer\def \table#1\par{\global \advance \tableno by 1%
\begingroup \closeup
\global\setbox\ListOfTables=\vbox{\unvbox\ListOfTables%
\smallskip\noindent{\bf\lasttableno.\ }{\strut\sl #1\strut}\par}\endgroup%
\medskip\caption{{\bf\lasttableno.\ }{\sl #1}}}%

\paperstyle

\newbox\myrefs%
    \setbox\myrefs=\vbox{%
\ifsastyle\noindent{\bigtype \bf{References}\medskip}
\else\smallskip\noindent\bf References\par\medskip\fi}%

\beginrefs 
\begingroup\raggedright%
\nohyphens%

\refarticle{refA11}
{\author{V.S.}{Afra\u\i movich}, \author{V.V.}{Bykov} \and
 \author{L.P.}{Shil'nikov}}{1983}
{On structurally unstable attracting limit sets of Lorenz attractor type}
            {Trans. Moscow Math. Soc.}{44}{153--216}
 
\refartbook{refA19}
{\author{V.S.}{Afra\u\i movich} \and \author{L.P.}{Shil'nikov}}{1983}
            {Strange attractors and quasiattractors}
            {Nonlinear Dynamics and Turbulence}
            {\author{G.I.}{Barenblatt}, \author{G.}{Iooss} \and
             \author{D.D.}{Joseph}}
            {Pitman}{Boston}{1--34}
 
\refarticle{refA18}
{\author{G.}{Ahlers} \and \author{I.}{Rehberg}}{1986}
            {Convection in a binary mixture heated from below}
            {Phys. Rev. Lett.}{56}{1373--1376}
 
\refarticle{refA21}
{\author{K.H.}{Alfsen} \and \author{J.}{Fr\o yland}}{1985}
            {Systematics of the Lorenz model at $\sigma=10$}
            {Physica Scripta}{31}{15--20}
 
\refarticle{refA22}
{\author{I.S.}{Aranson}, \author{V.S.}{Afra\u\i movich} \and
 \author{M.I.}{Rabinovich}}{1989}
            {Multidimensional strange attractors and turbulence}
            {Sov. Sci. Rev.~C. Math. Phys.}{8}{293--376}
 
\refarticle{refA10}
{\author{A.}{Arn\'eodo}, \author{P.}{Coullet} \and \author{C.}{Tresser}}{1981}
            {A possible new mechanism for the onset of turbulence}
            {Phys. Lett.}{81A}{197--201}
 
\refarticle{refA4}
{\author{A.}{Arn\'eodo}, \author{P.H.}{Coullet} \and \author{E.A.}{Spiegel}}{1982}
            {Chaos in a finite macroscopic system}
            {Phys. Lett.}{92A}{369--373}
 
\refarticle{refA2}
{\author{A.}{Arn\'eodo}, \author{P.H.}{Coullet} \and \author{E.A.}{Spiegel}}
            {1985}
            {The dynamics of triple convection}
            {Geophys. Astrophys. Fluid Dynamics}{31}{1--48}
 
\refarticle{refA3}
{\author{A.}{Arn\'eodo}, \author{P.H.}{Coullet}, \author{E.A.}{Spiegel} \and
 \author{C.}{Tresser}}
            {1985}
            {Asymptotic chaos}
            {Physica}{14D}{327--347}
 
\refarticle{refA13}
{\author{A.}{Arn\'eodo} \and \author{O.}{Thual}}{1985}
            {Direct numerical simulations of a triple convection problem
             versus normal form predictions}
            {Phys. Lett.}{109A}{367--373}
 
\refarticle{refA15}
{\author{V.I.}{Arnol'd}}{1972}
           {Lectures on bifurcations in versal families}
           {Russ. Math. Surveys}{27}{54--123}
 
\refbook{refA14}
{\author{V.I.}{Arnol'd}}{1983}
            {Geometrical Methods in the Theory of Ordinary Differential
             Equations}
            {Springer}{New York}
 
\refarticle{refA9}
{\author{W.}{Arter}}{1983}
            {Nonlinear convection in an imposed horizontal magnetic field}
            {Geophys. Astrophys. Fluid Dynamics}{25}{259--292}
 
\refbook{refA100}
{\author{L.}{Arnold}}{1974}
    {Stochastic Differential Equations}
    {Wiley}{New York}

\refbook{refA101}
{\author{Milton}{Abramowitz} \and \author{Irene}{Stegun}}{1968}
    {Handbook of Mathematical Functions}
    {Dover}{New York}
 
\refarticle{refA102}
{\author{*}{McA**}}{1888}
           {Lognormal stuff}
           {J.~Appl. Math. Phys. (ZAMP)}{37}{608--623}

\refarticle{refA103}
{\author{Guenter}{Ahlers}, \author{Christopher W.}{Meyer}
\& \author{I.}{Rehberg}}{1989}
            {Deterministic and Stochastic Effects Near the Convective Onset}
            {J. Stat. Phys.}{54}{1121--1131}
\refarticle{refA104}
{\author{Guenter}{Ahlers}, \author{M.C.}{Cross}, \author{P.C.}{Hohenberg}
\and \author{S.}{Safran}}{1981}
            {The amplitude equation near the convective threshold:
application to time-dependent heating experiments}
            {J. Fluid Mech.}{110}{297--334}

\refbook{refA105}
{\author{V.I.}{Arnol'd (Ed.)}}{1994}
            {Dynamical Systems V}
            {Springer}{Berlin}
\refbook{refA106}
{\author{Robert J.}{Adler}}{1981}
            {The Geometry of Random Fields}
            {Wiley}{Chichester}
\refbook{refA107}
{\author{Robert J.}{Adler}}{1990}
            {An Introduction to Continuity, Extrema and Related
Topics for General Gaussian Processes}
            {Institute of Mathematical Statistics}{Hayward, California}

\refarticle{refA108}
{\author{Dieter}{Armbruster}, \author{John}{Guckenheimer}
 \and \author{Philip}{Holmes}}{1988}
           {Heteroclinic cycles and modulated travelling waves in
systems with O(2) symmetry}
           {Physica D}{29}{257--282}
 
\refbook{refB2}
{\author{P.}{Berg\'e}, \author{Y.}{Pomeau} \and \author{C.}{Vidal}}{1984}
            {Order Within Chaos}
            {Wiley}{New York}

\refarticle{refB3}
{\author{C.S.}{Bretherton} \and \author{E.A.}{Spiegel}}{1983}
           {Intermittency through modulational instability}
           {Phys. Lett.}{96A}{152--156}
 
\refarticle{refB25}
{\author{F.H.}{Busse} \and \author{A.C.}{Or}}{1986}
           {Subharmonic and asymmetric convection rolls}
           {J.~Appl. Math. Phys. (ZAMP)}{37}{608--623}
 
\refarticle{refB100}
{\author{G.}{Broggi}, \author{A.}{Colombo}, \author{L.A.}{Lugiato}
\and \author{Paul}{Mandel}}{1986}
          {Influence of white noise on delayed bifurcations}
           {Phys. Rev. A}{33}{3635--3637}

\refarticle{refB101}
{\author{C.}{van den Broeck} \and \author{Paul}{Mandel}}{1987}
           {Delayed bifurcations in the presence of noise}
           {Phys. Lett.}{A122}{36--38}

\refarticle{refB102}
{\author{Jean-Marie}{Wersinger}, \author{John M.}{Finn}
  \and \author{Edward}{Ott}}{1980}
            {Bifurcations and strange behaviour in instability
saturation by nonlinear mode coupling}
         {Phys. Rev. Lett.}{44}{453--457}

\refarticle{refB103}
{\author{J.Robert}{Buchler}, \author{Pawe\l}{Moskalik} \and\author{G\'eza}{Kov\'acs}}{1991}
           {Periodic stellar pulsations: stability analysis and
amplitude equations}
           {Ap. J.}{380}{185--199}

\refarticle{refB104}
{\author{J.R.}{Buchler} \and\author{G.}{Kov\'acs}}{1986}
            {The effects of a 2:1 resonance in nonlinear radial
stellar pulsations}
            {Ap. J.}{303}{749--765}

\refartbook{refB105}
{\author{Claude}{Baesens}}{1991}
{Noise effect on dynamic bifurcations: the case of a
           period-doubling cascade}
            {Dynamic Bifurcations}
            {\author{E.}{Beno\^\i t}}
            {Springer}{Berlin}{pp107--130}

\refartbook{refB106}
{\author{E}{Beno\^\i t)}}{1991}
            {Linear dynamic bifurcation with noise}
            {Dynamic Bifurcations}
            {\author{E.}{Beno\^\i t}}
            {Springer}{Berlin}{pp131--150}

\refbook{refB107}
{\author{Denis R.}{Bell}}{1987}
            {The Malliavin Calculus}
            {Longman}{New York}
 
\refarticle{refB108}
{\author{A.R.}{Bulsara} \author{W.C.}{Schieve} \and
\author{E.W.}{Jacobs}}{1990}
            {Homoclinic chaos in a system perturbed by weak Langevin noise}
            {Phys. Rev. A}{41}{668--681}

\refarticle{refB109}
{\author{Per}{Bak}, \author{Chao}{Tang} \and
\author{Kurt}{Wiesenfeld}}{1988}
            {Self-organized criticality}
            {Phys. Rev. A}{38}{364--374}

\refbook{refB110}
{\author{Nicola}{Bellomo} \and \author{Rioccardo}{Riganti}}{1987}
            {Nonlinear Stochastic Systems in Physics and Mathematics}
            {World Scientific}{Singapore}
 
\refarticle{refB111}
{\author{S.M.}{Baer}, \author{T.}{Erneux}}{1986}
            {Singular Hopf bifurcation to relaxation oscillations}
            {SIAM J. Appl. Math.}{46}{721--739}

\refarticle{refB112}
{\author{S.M.}{Baer}, \author{T.}{Erneux}}{1992}
            {Singular Hopf bifurcation to relaxation oscillations II}
            {SIAM J. Appl. Math.}{52}{1651--1664}

\refarticle{refB113}
{\author{Claude}{Baesens}}{1991}
            {Slow sweep through a period-doubling cascade:
Delayed bifurcations and renormalisation}
            {Physica D}{53}{319--375}

\refarticle{refB114}
{\author{Andreas}{Becker} \and \author{Lorenz}{Kramer}}{1994}
            {Linear stability analysis for bifurcations in systems
with fluctuating control parameter}
            {Phys. Rev. Lett.}{73}{955--958}

\refarticle{refB115}
{\author{Paul}{Bryant} \and \author{Kurt}{Wiesenfeld}}{1986}
            {Suppression of period-doubling and nonlinear parametric
effects in periodically perturbed systems}
            {Phys. Rev. A}{33}{2525--2543}

\refbook{refB200}
{\author{E.}{Beno\^\i t (Ed.)}}{1991}
            {Dynamic bifurcations}
            {Springer}{Berlin}
 
\refbook{refC23}
{\author{J.}{Carr}}{1981}
    {Applications of Centre Manifold Theory}
    {Springer}{New York}
 
\refphd{refC30}
{\author{F.}{Cattaneo}}{1984}
            {Compressible magnetoconvection}
            {University Cambridge}
 
\refartbook{refC31}
{\author{F.}{Cattaneo}}{1984}
            {Oscillatory convection in sunspots}
            {The Hydromagnetics of the Sun}
            {\author{T.D.}{Guyenne} \and \author{J.J.}{Hunt}}
            {ESA}{Noordwijkerhoute}{47--50}
 
\refbook{refC22}
{\author{S.}{Chandrasekhar}}{1961}
            {Hydrodynamic and Hydromagnetic Stability}
            {Clarendon Press}{Oxford}
 
\refarticle{refC17}
{\author{J.}{Coste} \and \author{N.}{Peyraud}}{1982}
            {A new type of period-doubling bifurcations in one-dimensional
             transformations with two extrema}
            {Physica}{5D}{415--420}
 
\refarticle{refC3}
{\author{P.H.}{Coullet} \and \author{E.A.}{Spiegel}}{1983}
            {Amplitude equations for systems with competing instabilities}
            {SIAM J.~Appl. Math.}{43}{776--821}
 
\refarticle{refC15}
{\author{P.}{Coullet}, \author{J.-M.}{Gambaudo} \and \author{C.}{Tresser}}{1984}
            {Une nouvelle bifurcation de codimension~2: le collage de cycles}
            {C.~R.~Acad. Sc. Paris}{299}{S\'erie~I, 253--256}
 
\refarticle{refC7}
{\author{J.H.}{Curry}, \author{J.R.}{Herring},
 \author{J.}{Loncaric} \and \author{S.A.}{Orszag}}{1984}
            {Order and disorder in two- and three-dimensional B\'enard
             convection}
            {J.~Fluid Mech.}{147}{1--38}

\refbook{refC100}
{\author{P.}{Collet} \and \author{J.-P.}{Eckmann}}{1986}
    {Iterated maps of the interval as dynamical systems}
    {Birkh\ddot auser}{Boston}
 
\refarticle{refC101}
{\author{J.P.}{Crutchfield}, \author{J.D.}{Farmer}}{1984}
            {Fluctuations and simple chaotic dynamics}
            {Phys. Rep.}{92}{45--82}
\refarticle{refC102}
{\author{Alex}{Craik}}{1992}
            {Second-harmonic resonance in non-conservative systems}
            {Wave Motion}{15}{173--183}

\refphd{refC103}
{\author{Carston C.}{Chow}}{1985}
            {Spatiotemporal chaos on the nonlinear three wave interaction}
            {Massachusetts Institute of Technology}

\refbook{refC104}
{\author{C.}{Canuto et al}}{1988}
    {Spectral Methods in Fluid Dynamics}
    {Springer}{New York}

\refarticle{refC105}
{\author{Hugues}{Chat\'e} and \author{Paul}{Manneville}}{1987}
            {Transition to Turbulence via Spatiotemporal Intermittency}
            {Phys. Rev. Lett.}{58}{112--115}

\refarticle{refC106}
{\author{M.C.}{Cross} \and \author{P.C.}{Hohenberg}}{1993}
            {Pattern formation outside of equilibrium}
            {Rev. Mod. Phys.}{92}{851--1089}

\refarticle{refC107}
{\author{M.}{Ciofini}, \author{R.}{Meucci}
\and \author{F.T.}{Arecchi}}{1990}
            {Transient statistics in a $CO_2$ laser with a slowly
swept pump}
            {Phys. Rev. A}{42}{482--486}

\refarticle{refC108}
{\author{Jack}{Carr} \and \author{Robert}{Pego}}{1992}
            {Self-similarity in a coarsening model in one dimension}
            {Proc. Royal Soc. London A}{436}{569--582}

\refarticle{refC109}
{\author{J.}{Carr} \and \author{R.L.}{Pego}}{1989}
            {Metastable Patterns in Solutions of $u_t=\ep^2u_{xx}-f(u)$}
            {Commun. Pure Appl. Math.}{42}{523--576}

\refarticle{refC111}
{\author{Jack}{Carr} \and \author{Robert}{Pego}}{1992}
            {Invariant manifolds for metastable patterns in
        $u_t=\ep^2u_{xx}-f(u)$}
            {Proc. Royal Soc. Edinburgh A}{116}{133--160}
\refbook{refC110}
{\author{H.S.}{Carslaw} \and \author{J.C.}{Jaeger}}{1959}
    {Conduction of heat in solids}
    {Oxford University Press}{London}

\refarticle{refC111}
{\author{J.P.}{Crutchfield}, \author{B.A.}{Huberman}}{1980}
            {Fluctuations and the onset of chaos}
            {Phys. Lett. A}{77}{407--410}

\refarticle{refC112}
{\author{Thomas}{Clune}, \author{Edgar}{Knobloch}}{1994}
            {pattern selection in three-dimensional magnetoconvection}
            {Physica D}{74}{151--176}

\refarticle{refD2}
{\author{L.N.}{Da~Costa}, \author{E.}{Knobloch} \and \author{N.O.}{Weiss}}{1981}
            {Oscillations in double-diffusive convection}
            {J.~Fluid Mech.}{109}{25--43}
 
\refarticle{refD14}
{\author{G.}{Dangelmayr}, \author{D.}{Armbruster} \and
 \author{M.}{Neveling}}{1985}
            {A codimension three bifurcation for the laser with saturable
             absorber}
            {Z.~Phys.~B -- Condensed Matter}{59}{365--370}
 
\refarticle{refD3}
{\author{G.}{Dangelmayr} \and \author{E.}{Knobloch}}{1986}
            {Interaction between standing and travelling waves and steady
             states in magnetoconvection}
            {Phys. Lett.}{117A}{394--398}
 
\refarticle{refD1}
{\author{G.}{Dangelmayr} \and \author{E.}{Knobloch}}{1987}
            {The Takens--Bogdanov bifurcation with $O(2)$-symmetry}
            {Phil. Trans. R.~Soc. Lond.}{A322}{243--279}
 
\refarticle{refD6}
{\author{A.E.}{Deane}, \author{E.}{Knobloch} \and \author{J.}{Toomre}}{1988}
            {Travelling waves in large-aspect-ratio thermosolutal convection}
            {Phys. Rev.}{37A}{1817--1820}
 
\refarticle{refD11}
{\author{A.E.}{Deane} \and \author{L.}{Sirovich}}{1991}
            {A computational study of Rayleigh--B\'enard convection. Part~1.
             Rayleigh-number scaling}
            {J.~Fluid Mech.}{222}{231--250}
 
\refarticle{refD15}
{\author{P.}{Deuflhard}}{1983}
            {Order and stepsize control in extrapolation methods}
            {Numer. Math.}{41}{399--422}
 
\refbook{refD13}
{\author{E.}{Doedel}}{1986}%
            {AUTO: Software for Continuation and Bifurcation Problems in
             Ordinary Differential Equations}
            {C.I.T. Press}{Pasadena}

\refarticle{refD100}
{\author{M.I.}{Dykman},
\author{R.}{Mannella}, \author{P.V.E.}{McClintock},
\author{N.D.}{Stein} \and \author{N.G.}{Stocks}}{1993}
           {Probability distributions and escape rates for systems
driven by quasimonochromatic noise}
           {Phys. Rev. E}{47}{3996--4009}

\refarticle{refD101}
{\author{I.T.}{Drummond}}{1993}
           {Multiplicative stochastic differential equations with
noise-induced transitions}
           {J. Phys. A}{25}{2273--2296}

\refarticle{refD102}
{\author{S.M.}{Catterall}, \author{I.T.}{Drummond}
\and \author{R.R.}{Horgan}}{1991}
           {Stochastic simulation of quantum mechanics}
           {J. Phys. A}{24}{4081--4091}

\refarticle{refD103}
{\author{Didier}{Dangoisse}, \author{Pierre}{Glorieux}
\and \author{Daniel}{Hennequin}}{1987}
           {Chaos in a $CO_2$ laser with modulated control parameters:
        Experiments and numerical simulations}
           {Phys. Rev. A}{36}{4775--4791}

\refarticle{refD104}
{\author{Charles R.}{Doering}}{1987}
           {Nonlinear Parabolic Stochastic Differential Equations
with Additive Colored Noise on $\real^d\times\real_+$:
A Regulated Stochastic Quantization}
           {Commun. Math. Phys.}{109}{537--561}

\refarticle{refD105}
{\author{Charles R.}{Doering}}{1987}
           {A stochastic partial differential equation
with multiplicative noise}
           {Phys. Lett. A}{122}{133--139}

\refarticle{refD106}
{\author{Marc}{Diener}}{1984}
           {The Canard Unchained or How Fast/Slow Dynamical Systems Bifurcate}
           {The Mathematical Intelligencer}{6}{No.2, 38--49}

\refartbook{refD107}
{\author{C.R.}{Doering}}{1992}
            {Evaluating the Path Integral:
Field Theory Estimates via Stochastic Quantization}
            {Lectures on Path Integration}
            {\author{R.}{Cerdeira}}
            {World Scientific}{Singapore}{pp184--214}

\refartbook{refD108}
{\author{Marc}{Diener}}{1981}
            {On the Perfect Delay Convention, or
the Revolt of the Slaved Variables}
            {Chaos and Order in Nature}
            {\author{H.}{Haken}}
    {Springer}{Berlin}{pp249--268}

\refartbook{refD109}
{\author{B.}{Candelpergher}, \author{F.}{Diener} \and
\author{M.}{Diener}}{1990}
{Retard \`a la bifurcation: du local au global}
            {Bifurcations of planar vector fields}
            {\author{J.P.}{Francoise} \and \author{R.}{Roussarie}}
            {Springer}{Berlin}{pp1--19}

\refarticle{refD110}
{\author{E.}{Beno\^it}, \author{F.}{Diener} \and
\author{M.}{Diener}}{1981}
{Chasse au canard}
            {Collectanea Mathematica}
            {31}{37--119}

\refbook{refD111}
{\author{P.}{Drazin}}{1992}
    {Nonlinear Systems}
    {Cambridge University Press}{Cambridge}

\refarticle{refE100}
{\author{A.}{Einstein}}{1905}
           {\"uber die von der molekular-kinetischen Theorie der
        W\"arme geforderte Bewegung von in ruhended Fl\"ussigkeiten}
           {Ann.Phys.}{17}{549-555}

\refartbook{refE101}
{\author{T.}{Erneux}, \author{E.C.}{Reiss}, \author{L.J.}{Holden}
\and \author{M.}{Georgiou}}{1991}
            {Slow Passage through bifurcation and limit points.
Asymptotic theory}
            {Dynamic Bifurcations}
            {\author{E.}{Beno\^\i t}}
            {Springer}{Berlin}{pp14--28}
 
\refarticletobe{refE102}
{\author{Stephen}{Einchcomb} \and \author{A.J.}{McKane}}{1993}
           {Quasi-monochromatic noise stuff with path integrals}
           {Phys. Rev. E}

\refartbook{refE103}
{\author{W.}{Eckhaus}}{1983}
{Relaxation oscillations including a standard chase on french ducks}
            {Asymptotic Analysis}
            {\author{F.}{Verhulst}}
            {Springer}{Berlin}{pp432--449}

\refarticle{refE104}
{\author{T.}{Erneux} \and \author{Paul}{Mandel}}{1984}
            {Slow passage through laser first threshold}
            {Phys. Rev. A}{39}{5179--5188}

\refarticle{refF5}
{\author{A.C.}{Fowler}, \author{J.D.}{Gibbon} \and
 \author{M.J.}{McGuinness}}{1983}
            {The real and complex Lorenz equations and their relevance to
             physical systems}
            {Physica}{7D}{126--134}
 
\refarticle{refF4}
{\author{A.C.}{Fowler}}{1990}
            {Homoclinic bifurcations in $n$ dimensions}
            {Studies in Applied Mathematics}{83}{193--209}

\refarticle{refF100}
{\author{L.}{Fronzoni}, \author{Frank}{Moss} \and
\author{P.V.E.}{McClintock}}{1987}
{Swept-parameter-induced postponements and noise on the Hopf bifurcation}
            {Phys. Rev. A}{36}{1492--1494}

\refarticle{refF101}
{\author{K.}{Fujimura} \and \author{R.E.}{Kelly}}{1994}
{Interaction between longitudinal convection rolls and transverse
waves in unstably stratified plane Poiseuille flow}
            {Phys. Fluids}{7}{25--65}

\refarticle{refF102}
{\author{Ronald Forrest}{Fox}}{1972}
{Contributions to the Theory of Multiplicative Stochastic Processes}
            {J. Math. Phys}{13}{1196--1207}
 
\refarticle{refF103}
{\author{Tadahisa}{Funaki}}{1983}
{Random motion of strings and related stochastic evolution equations}
            {Nagoya Math. J.}{89}{129--193}

\refarticle{refF104}
{\author{Tadahisa}{Funaki}}{1995}
{The scaling limit for a stochastic PDE and the separation of phases}
            {Probab. Theory Relat. Fields}{102}{221--288}

\refarticle{refG32}
{\author{P.}{Gaspard}, \author{R.}{Kapral} \and \author{G.}{Nicolis}}{1984}
            {Bifurcation phenomena near homoclinic systems: a two-parameter
             analysis}
            {J.~Stat. Phys.}{35}{697--727}

\refarticle{refG12}
{\author{P.}{Glendinning} \and \author{C.}{Sparrow}}{1984}
            {Local and global behaviour near homoclinic orbits}
            {J.~Stat. Phys.}{35}{645--696}
 
\refphd{refG22}
{\author{P.A.}{Glendinning}}{1985}
            {Homoclinic bifurcations}
            {University of Cambridge}
 
\refarticle{refG2}
{\author{P.}{Glendinning} \and \author{C.}{Sparrow}}{1986}
            {T-points: a codimension two heteroclinic bifurcation}
            {J.~Stat. Phys.}{43}{479--488}
 
\refarticle{refG45}
{\author{I.}{Goldhirsch}, \author{R.B.}{Pelz}, \author{S.A.}{Orszag}}{1989}
            {Numerical simulation of thermal convection in a two-dimensional
             finite box}
            {J.~Fluid Mech.}{199}{1--28}
 
\refbook{refG31}
{\author{D.}{Gottlieb} \and \author{S.A.}{Orszag}}{1977}
    {Numerical Analysis of Spectral Methods: Theory and Applications}
    {Society for Industrial and Applied Mathematics}{Philadelphia}
 
\refarticle{refG41}
{\author{E.}{Graham}}{1975}
            {Numerical simulation of two-dimensional compressible convection}
            {J.~Fluid Mech.}{70}{689--703}
 
\refarticle{refG42}
{\author{R.}{Graham}}{1976}
            {Onset of self-pulsing in lasers and the Lorenz model}
            {Phys. Lett.}{58A}{440--442}
 
\refarticle{refG43}
{\author{S.}{De~Gregorio}, \author{E.}{Scoppola} \and
 \author{B.}{Tirozzi}}{1983}
            {A rigorous study of periodic orbits by means of a computer}
            {J.~Stat. Phys.}{32}{25--33}
 
\refartbook{refG28}
{\author{J.}{Guckenheimer}}{1976}
            {A strange, strange attractor}
            {The Hopf Bifurcation and Its Applications}
            {\author{J.E.}{Marsden} \and \author{M.}{McCracken}}
            {Springer}{New York}{368--381}
 
\refarticle{refG30}
{\author{J.}{Guckenheimer} \and \author{E.}{Knobloch}}{1983}
            {Nonlinear convection in a rotating layer: amplitude expansions and
             normal forms}
            {Geophys. Astrophys. Fluid Dynamics}{23}{247--272}
 
\refbook{refG1}
{\author{J.}{Guckenheimer} \and \author{P.}{Holmes}}{1987}
            {Nonlinear Oscillations,
             Dynamical Systems and Bifurcations of Vector Fields}
            {Springer}{New York}

\refbook{refG100}
{\author{C.W.}{Gardiner}}{1990}
    {Handbook of Stochastic Methods}
    {Springer}{Berlin}

\refbook{refG101}
{\author{I.S.}{Gradshteyn} \and \author{M.}{Ryzhik}}{1980}
            {Table of integrals, series, and products} 
            {Academic Press}{New York}

\refartbook{refG102}
{\author{Paolo}{Grigolini}}{1989}
            {The projection approach to the Fokker-Planck equation:
application to phenomenological stochastic equations with colored noise}
            {Noise in nonlinear dynamical systems, Vol. I}
            {\author{Frank}{Moss} \and \author{P.V.E.}{McClintock}}
            {Cambridge University Press}{Cambridge}{pp161--190}

\refartbook{refG103}
{\author{Robert}{Graham}}{1989}
            {Macroscopic potentials, bifurcations and noise in
dissipative systems}
            {Noise in nonlinear dynamical systems, Vol. I}
            {\author{Frank}{Moss} \and \author{P.V.E.}{McClintock}}
            {Cambridge University Press}{Cambridge}{225--278}
  
\refarticle{refG104}
{\author{I.}{Gy\"ongy} \and \author{E.}{Pardoux}}{1993}
            {On quasi-linear stochastic partial differential equations}
            {Probab. Theory Relat. Fields}{94}{413--426}

\refarticle{refG105}
{\author{I.}{Gy\"ongy} \and \author{E.}{Pardoux}}{1993}
            {On the regularization effect of space-time white noise on
quasi-linear parabolic partial differential equations}
            {Probab. Theory Relat. Fields}{97}{211--230}

\refarticle{refG106}
{\author{R.}{Graham} \and \author{T.}{T\'el}}{1990}
            {Steady-state ensemble for the complex Ginzburg-Landau
equation with weak noise}
            {Phys. Rev. A}{42}{4661--4677}

\refarticle{refG107}
{\author{Hermann}{Grabert} \author{Peter}{H\ddot anggi}
\and \author{Peter}{Talkner}}{1979}
           {Is quantum mechanics equivalent to a classical stochastic process?}
            {Phys. Rev. A}{19}{2440--2445}

\refarticle{refG108}
{\author{J.}{Garc\'\i a-Ojalvo}
\and \author{J.M.}{Sancho}}{1994}
           {Colored noise in spatially extended systems}
            {Phys. Rev. E}{49}{2769--2778}

\refarticle{refG109}
{\author{J.}{Garc\'\i a-Ojalvo}, \author{J.M.}{Sancho}
\and \author{L.}{Ram\'\i rez-Piscina}}{1992}
           {Generation of spatiotemporal colored noise}
            {Phys. Rev. A}{46}{4670--4675}

\refarticle{refG110}
{\author{Robert}{Graham}}{1974}
            {Hydrodynamic fluctuations near the convection instability}
            {Phys. Rev. A}{10}{1762--1784}
\refarticle{refG111}
{\author{R.}{Graham} \and \author{A.}{Schenzle}}{1982}
            {Stabilization by multiplicative noise}
            {Phys. Rev. A}{26}{1676--1685}

\refarticle{refG112}
{\author{J.}{Garc\'\i a-Ojalvo}, \author{A.}{Hern\'andez-Machado} \and
\author{J.M.}{Sancho}}{1993} 
           {Effects of External Noise on the Swift-Hohenberg equation}
            {Phys. Rev. Lett.}{71}{1542--1545}

\refarticle{refG113}
{\author{P.}{Gaspard} et al (Ed.)}{1991}
            {Homoclinic Chaos}
            {Physica D}{62}{1--372}

\refarticle{refG114}
{\author{Raymind}{Goldstein} \and
\author{Gemunu H.}{Gunaratne}}{1991} 
           {Hydrodynamic and interfacial patterns with broken
space-time symmetry}
            {Phys. Rev. A}{43}{6700--6710}

\refarticle{refH13}
{\author{H.}{Haken}}{1975}
            {Analogy between higher instabilities in fluids and lasers}
            {Phys . Lett.}{53A}{77--78}
 
\refarticle{refH17}
{\author{S.M.}{Hammel}, \author{J.A.}{Yorke} \and \author{C.}{Grebogi}}{1988}
            {Numerical orbits of chaotic processes represent true orbits}
            {Bull. Am. Math. Soc.}{19}{465--469}
 
\refbook{refH15}
{\author{A.C.}{Hearne}}{1987}
            {Reduce User's Manual -- Version 3.3}
            {Rand}{Santa Monica}
 
\refarticle{refH3}
{\author{D.W.}{Hughes} \and \author{M.R.E.}{Proctor}}{1988}
            {Magnetic fields in the solar convection zone: magnetoconvection
             and magnetic buoyancy}
            {Ann. Rev. Fluid Mech.}{20}{187--223}
 
\refarticle{refH5}
{\author{H.E.}{Huppert} \and \author{D.R.}{Moore}}{1976}
            {Nonlinear double-diffusive convection}
            {J.~Fluid Mech.}{78}{821--854}
 
\refarticle{refH16}
{\author{N.E.}{Hurlburt} \and \author{J.}{Toomre}}{1988}
            {Magnetic fields interacting with nonlinear compressible convection}
            {Ap. J.}{327}{920--932}
 
\refarticle{refH1}
{\author{N.E.}{Hurlburt}, \author{M.R.E.}{Proctor},
 \author{N.O.}{Weiss} \and \author{D.P.}{Brownjohn}}{1989}
            {Nonlinear compressible magnetoconvection.
             Part~1. Travelling waves and oscillations}
            {J.~Fluid Mech.}{207}{587--628}

\refbook{refH100}
{\author{H.}{Haken(ed)}}{1981}
    {Chaos and Order in Nature}
    {Springer}{Berlin}
 
\refarticle{refH101}
{\author{D.W.}{Hughes} \and \author{M.R.E.}{Proctor}}{1990}
            {Chaos and the effect of noise in a model of three-wave
mode coupling}
            {Physica D}{46}{163--176}

\refarticle{refH102}
{\author{D.W.}{Hughes} \and \author{M.R.E.}{Proctor}}{1990}
            {A low order model of the shear instability of convection:
chaos and the effect of noise}
            {Nonlinearity}{3}{127--153}

\refarticle{refH103}
{\author{M.R.E.}{Proctor} \and\author{D.W.}{Hughes}}{1990}%
{The false Hopf bifurcation and noise sensitivity
        in bifurcations with symmetry}
            {Eur.J.Mech.,B/Fluids}{10}{81--86}

\refarticle{refH104}
{\author{Rebecca L.}{Honeycutt}}{1992}
            {Stochastic Runge-Kutta algorithms. I. White noise}
            {Phys. Rev. A}{45}{600--603}

\refbook{refH105}
{\author{Harry}{Hochstadt}}{1971}
    {The functions of mathematical physics}
    {Dover}{New York}

\refartbook{refH106}
{\author{Peter}{Hanggi}}{1989}
            {Colored noise in continuous dynamical systems: a
functional calculus approach}
            {Noise in nonlinear dynamical systems, Vol. I}
            {\author{Frank}{Moss} \and \author{P.V.E.}{McClintock}}
            {Cambridge University Press}{Cambridge}{307--328}
 
\refbook{refH107}
{\author{H.}{Haken}}{1978}
    {Synergetics}
    {Springer}{Berlin}

\refarticle{refH108}
{\author{Richard}{Haberman}}{1979}
            {Slowly varying jump and transition phenomena associated
with algebraic bifurcation problems}
            {SIAM J. Appl. Math.}{37}{69--106}

\refarticle{refH109}
{\author{J.J.}{Healy}, \author{D.S.}{Broomhead}, \author{K.A.}{Cliff},
\author{R.}{Jones} \and \author{T.}{Mullin}}{1991}
            {The origins of chaos in a modified Van der Pol oscillator}
            {Physica D}{48}{322--339}

\refbook{refH110}
{\author{W.}{Horsthemke} \& \author{R.}{Lefever}}{1984}
    {Noise-induced transitions}
    {Springer}{Berlin}
 
\refarticle{refH111}
{\author{W.}{Horsthemke} \and \author{R.}{Lefever}}{1977}
    {Phase transitions induced by external noise}
        {Phys. Lett. A}{64}{19--21}
 
\refarticle{refH112}
{\author{Michael R. E.}{Proctor} \and \author{Judith Y.}{Holyer}}{1986}
    {Planform selection in salt fingers}
        {J. Fluid Mech.}{168}{241--253}
\refarticle{refH113}
{\author{Lisa}{Holden} \and \author{Thomas}{Erneux}}{1993}
    {Slow passage through a Hopf bifurcation:
from oscillatory to steady state solutions}
        {SIAM J. Appl. Math.}{53}{1045--1058}
 
\refarticle{refH114}
{\author{P.}{Holmes}}{1979}
    {A nonlinear oscillator with a strange attractor}
        {Phil. Trans.}{292}{419--448}
\refarticle{refH115}
{\author{P.}{Holmes}}{1990}
    {Poincar\'e, celestial mechanics, dynamical-systems theory and `chaos'}
        {Phys. Rep.}{193}{137--163}
 
\refarticle{refH116}
{\author{L.N.}{Howard} \and \author{R.}{Krishnamurti}}{1986}
    {Large-scale flow in turbulent convection:
a mathematical model}
        {J. Fluid Mech.}{170}{385--410}

\refarticle{refH117}
{\author{P.C.}{Hohenberg} \and \author{J.W.}{Swift}}{1992}
            {Effects of additive noise at the onset of 
Rayleigh-B\'enard convection}
            {Phys. Rev. A}{46}{4773--4785}

\refarticle{refI100}
{\author{Kiyosi}{Ito}}{1964}
            {Expected number of zeros of continuous stationary Gaussian processes}
            {J. Math. Kyoto Univ.}{3-2}{207--216}
 
\refarticle{refJ100}
{\author{Peter}{Jung}, \author{George}{Gray}, \author{Rajarshi}{Roy}
 \and \author{Paul}{Mandel}}{1990}
    {Scaling Law for Dynamical Hysteresis}
        {Phys. Rev. Lett.}{65}{1873--1876}

\refphd{refJ5}
{\author{K.A.}{Julien}}{1991}
            {Strong spatial resonance in convection}
            {University of Cambridge}
 
\refarticle{refK2}
{\author{E.}{Knobloch} \and \author{M.R.E.}{Proctor}}{1981}
            {Nonlinear periodic convection in double-diffusive systems}
            {J.~Fluid Mech.}{108}{291--316}
 
\refarticle{refK5}
{\author{E.}{Knobloch}, \author{N.O.}{Weiss} \and \author{L.N.}{Da~Costa}}{1981}
            {Oscillatory and steady convection in a magnetic field}
            {J.~Fluid Mech.}{113}{153--186}
 
\refarticle{refK7}
{\author{E.}{Knobloch} \and \author{N.O.}{Weiss}}{1983}
            {Bifurcations in a model of magnetoconvection}
            {Physica}{9D}{379--407}
 
\refarticle{refK11}
{\author{E.}{Knobloch} \and \author{N.O.}{Weiss}}{1984}
            {Convection in sunspots and the origin of umbral dots}
            {Mon.~Not. R.~Astr. Soc.}{207}{203--214}
 
\refarticle{refK8}
{\author{E.}{Knobloch}}{1986}
            {On convection in a horizontal magnetic field with periodic
             boundary conditions}
            {Geophys. Astrophys. Fluid Dynamics}{36}{161--177}
 
\refarticle{refK19}
{\author{E.}{Knobloch}}{1986}
            {On the degenerate Hopf bifurcation with $O(2)$ symmetry}
            {Contemp. Math.}{56}{193--201}
 
\refarticle{refK24}
{\author{E.}{Knobloch}}{1986}
            {Normal forms for bifurcations at a double zero eigenvalue}
            {Phys. Lett.}{115A}{199--201}
 
\refarticle{refK8a}
{\author{E.}{Knobloch}}{1986a}
            {On convection in a horizontal magnetic field with periodic
             boundary conditions}
            {Geophys. Astrophys. Fluid Dynamics}{36}{161--177}
 
\refarticle{refK19b}
{\author{E.}{Knobloch}}{1986b}
            {On the degenerate Hopf bifurcation with $O(2)$ symmetry}
            {Contemp. Math.}{56}{193--201}
 
\refarticle{refK24c}
{\author{E.}{Knobloch}}{1986c}
            {Normal forms for bifurcations at a double zero eigenvalue}
            {Phys. Lett.}{115A}{199--201}
 
\refarticle{refK4}
{\author{E.}{Knobloch}, \author{D.R.}{Moore},
 \author{J.}{Toomre} \and \author{N.O.}{Weiss}}{1986}
            {Transitions to chaos in two-dimensional double-diffusive
             convection}
            {J.~Fluid Mech.}{166}{409--448}
 
\refarticle{refK1}
{\author{E.}{Knobloch} \and \author{M.R.E.}{Proctor}}{1988}
            {The double Hopf bifurcation with $2:1$ resonance}
            {Proc. R.~Soc. Lond.}{415A}{61--90}
 
\refarticle{refK17}
{\author{E.}{Knobloch} \and \author{M.}{Silber}}{1990}
            {Travelling wave convection in a rotating layer}
            {Geophys. Astrophys. Fluid Dynamics}{51}{195--209}

\refbook{refK34}
{\author{D.E.}{Knuth}}{1984}
            {The \TeX book}
            {Addison Wesley}{Reading}
\refbook{refK100}
{\author{Peter E.}{Kloeden} \and \author{Eckhard}{Platen}}{1992}
    {Numerical Solution of Stochastic Differential Equations}
    {Springer}{Berlin}

\refbook{refK101}
{\author{H.}{Kunita}}{1990}
    {Stochastic Flows and Stochastic Differential Equations}
    {Cambridge University Press}{Cambridge}

\refbook{refK102}
{\author{T.W.}{K\ddot orner}}{1988}
    {Fourier analysis}
    {Cambridge University Press}{Cambridge}

\refarticle{refK103}
{\author{N.}{van Kampen}}{1981}
            {\ito\ Versus Stratonovich}
            {J. Stat. Phys.}{24}{175--187}
\refarticle{refK104}
{\author{D.K.}{Kondepudi} \and \author{G.W.}{Nelson}}{1985}
            {Weak neutral currents and the origin of biomolecular chirality}
            {Nature}{314}{438--441}
\refarticle{refK105}
{\author{D.K.}{Kondepudi}, \author{F.}{Moss} \and 
\author{P.V.E.}{McClintock}}{1986}
            {Observation of symmetry breaking, state selection and
sensitivity in a noisy electronic system}
            {Physica D}{21}{296--306}
 
\refbook{refK106}
{\author{Frank B.}{Knight}}{1981}
    {Essentials of Brownian motion and diffusion}
    {American Mathematical Society}{Providence}

\refarticle{refK107}
{\author{Raymond}{Kapral} \and \author{Paul}{Mandel}}{1985}
            {Bifurcation structure of the nonautonomous quadratic map}
            {Phys. Rev. A}{32}{1076--1081}

\refbook{refK108}
{\author{J.}{Kevorkian} \and \author{J.D.}{Cole}}{1981}
    {Perturbation Methods In Applied Mathematics}
    {Springer}{New York}

\refarticle{refK109}
{\author{D.J.}{Kaup}, \author{A.}{Rieman} \and \author{A.}{Bers}}{1979}
            {Space-time evolution of nonlinear three-wave interactions}
            {Rev. Mod. Phys.}{51}{275--309}
 
\refbook{refK110}
{\author{Ioannis}{Karatzas} \and \author{Steven E.}{Shreve}}{1988}
    {Brownian Motion and Stochastic Calculus}
    {Springer}{New York}

\refarticle{refL5}
{\author{E.N.}{Lorenz}}{1963}
            {Deterministic nonperiodic flow}
            {J.~Atmos. Sci.}{20}{130--141}
 
\refarticle{refL12}
{\author{D.V.}{Lyubimov} \and \author{M.A.}{Zaks}}{1983}
            {Two mechanisms of the transition to chaos in finite-dimensional
             models of convection}
            {Physica}{9D}{52--64}
 
\refarticle{refL18}
{\author{D.V.}{Lyubimov}, \author{A.S.}{Pikovsky} \and
 \author{M.A.}{Zaks}}{1989}
            {Universal scenarios of transition to chaos via homoclinic
             bifurcations}
            {Sov. Sci. Rev.~C. Math. Phys.}{8}{221--292}

\refarticle{refL23}
{\author{A.S.}{Landsberg} \and \author{E.}{Knobloch}}{1991}
            {Direction-reversing traveling waves}
            {Phys. Lett. A}{159}{17--20}

\refarticle{refL24}
{\author{A.S.}{Landsberg} \and \author{E.}{Knobloch}}{1993}
            {New types of waves in systems with $O(2)$ symmetry}
            {Phys. Lett. A}{179}{316--324}

\refarticle{refL100}
{\author{G.D.}{Lythe} \and \author{M.R.E.}{Proctor}}{1993}
            {Noise and slow-fast dynamics in a three-wave resonance problem}
            {Phys. Rev. E. }{47}{3122-3127}
 
\refarticle{refL101}
{\author{M.R.E.}{Proctor} \and \author{G.D.}{Lythe}}{1993}
            {Noise and resonant mode interactions}
            {Ann. New York Acad. Sci. }{706}{42-53}

\refartbook{refL102}
{\author{Claude}{Lobry}}{1991}
            {Dynamic Bifurcations}
            {Dynamic Bifurcations}
            {\author{E.}{Beno\^\i t}}
            {Springer}{Berlin}{pp1--13}
 
\refartbook{refL103}
{\author{Katya}{Lindenberg}, {Bruce J.}{West} \and {Jaume}{Masoliver}}{1989}
            {First passage time problems for non-Markovian processes}
            {Noise in nonlinear dynamical systems, Vol. I}
            {\author{Frank}{Moss} \and \author{P.V.E.}{McClintock}}
            {Cambridge University Press}{Cambridge}{110--160}
 
\refbook{refL104}
{\author{L.}{Landau} \and \author{E.}{Lifshitz}}{1959}
    {Statistical Physics}
    {Pergamon}{London}

\refarticle{refL105}
{\author{N.R.}{Lebovitz} \and \author{R.J.}{Schaar}}{1975}
            {Exchange of stabilities in autonomous systems}
            {Studies in Appl. Math. }{54}{229--260}

\refarticle{refL106}
{\author{Claude}{Lobry}}{1992}
        {A propos du sens des textes math\'ematiques. Un example:
        la th\'eorie des ``bifurcations dynamiques''}
            {Ann. Inst. Fourier}{42}{327-352}

 
\refartbook{refL109}
{\author{G.D.}{Lythe}}{1994}
            {A noise-controlled dynamic bifurcation}
            {Chaos and Nonlinear Mechanics}
                {\author{T.}{Kapitaniak} and \author{J.}{Brindley}(eds)}
{World Scientific}{Singapore}{147--158}
 
\refbook{refL110}
{\author{Andrej}{Lasota} \and \author{Michael C.}{Mackey}}{1985}
    {Probabilistic properties of deterministic systems}
    {Cambridge University Press}{Cambridge}

\refarticle{refL111}
{\author{Pui-Man}{Lam} \and \author{Diola}{Bagayoko}}{1993}
            {Spatiotemporal correlation of coloured noise}
            {Phys. Rev. E}{48}{3267--3270}
\refarticle{refL112}
{\author{Katya}{Lindenberg}, {Bruce J.}{West} \and {Raoul}{Kopelman}}{1989}
            {Steady-State Segregation in Diffusion-Limited Reactions}
            {Phys. Rev. Lett.}{60}{1777--1780}

\refbook{refL113}
{\author{Leon}{Lapidus} \and \author{George F.}{Pinder}}{1982}
    {Numerical Solution of Partial Differential Equations in Science and Engineering}
    {Wiley}{New York}

\refartbook{refL114}
{\author{G.D.}{Lythe}}{1994}
            {Noise and Dynamic transitions}
            {Stochastic Partial Differential Equations}
                {\author{Alison}{Etheridge}}{Cambridge University Press}
                {Cambridge}{}

\refarticle{refL115}
{\author{G.D.}{Lythe}}{1995}
           {Dynamics controlled by additive noise}
           {Nuovo Cimento D}{17}{855--861}

\refphd{refL116}
{\author{G.D.}{Lythe}}{1994}
            {Stochastic slow-fast dynamics}
            {University of Cambridge}

\refarticletobe{refL117}
{\author{C.-H.}{Lu} \and \author{R.M.}{Evan-Iwanowski}}{1994}
           {The Nonstationary Effects on the Logistic Map and the
Softening Duffing Oscillator}
           {}

\refartbook{refM28}
{\author{B.A.}{Malomed} \and \author{A.A.}{Nepomnyashchy}}{1990}
            {Onset of chaos in the generalized Ginzburg--Landau equation}
            {Nonlinear Evolution of Spatio-Temporal Structures in Dissipative
             Continuous Systems}
            {\author{F.H.}{Busse} \and \author{L.}{Kramer}}
            {Plenum Press}{New York}{419--424}

\refarticle{refM25}
{\author{W.V.R.}{Malkus} \and \author{G.}{Veronis}}{1958}
            {Finite amplitude cellular convection}
            {J.~Fluid Mech.}{4}{225--260}
 
\refarticle{refM19}
{\author{P.S.}{Marcus}}{1981}
            {Effects of truncation in modal representations of thermal
             convection}
            {J.~Fluid Mech.}{103}{241--255}
 
\refartbook{refM28}
{\author{B.A.}{Malomed} \and \author{A.A.}{Nepomnyashchy}}{1990}
            {Onset of chaos in the generalized Ginzburg--Landau equation}
            {Nonlinear Evolution of Spatio-Temporal Structures in Dissipative
             Continuous Systems}
            {\author{F.H.}{Busse} \and \author{L.}{Kramer}}
            {Plenum Press}{New York}{419--424}
 
\refbook{refM34}
{\author{P.}{Manneville}}{1990}
            {Dissipative Structures and Weak Turbulence}
            {Academic Press}{Boston}
 
\refarticle{refM5}
{\author{R.M.}{May}}{1976}
            {Simple mathematical models with very complicated dynamics}
            {Nature}{261}{459--467}
 
\refarticle{refM20}
{\author{D.R.}{Moore} \and \author{N.O.}{Weiss}}{1973}
            {Two-dimensional Rayleigh--B\'enard convection}
            {J.~Fluid Mech.}{58}{289--312}
 
\refarticle{refM22}
{\author{D.R.}{Moore}, \author{N.O.}{Weiss} \and \author{J.M.}{Wilkins}}{1990}
            {The reliability of numerical experiments:
             transitions to chaos in thermosolutal convection}
            {Nonlinearity}{3}{997--1014}

\refarticle{refM100}
{\author{Paul}{Mandel} \and \author{T.}{Erneux}}{1984}
            {Laser Lorenz equations with a time-dependent parameter}
            {Phys. Rev. Lett.}{53}{1818--1820}

\refarticle{refM101}
{\author{Paul}{Mandel} \and \author{Thomas}{Erneux}}{1987}
            {The slow passage through a steady bifurcation:
        delay and memory effects}
            {J. Stat. Phys.}{48}{1059--1070}

\refarticle{refM102}
{\author{R.}{Mannella}, \author{Frank}{Moss}
\and \author{P.V.E.}{McClintock}}{1987}
           {Postponed bifurcations of a ring-laser with a swept parameter
        and additive colored noise}
           {Phys. Rev. A}{35}{2560--2566}
\refbook{refM103}
{\author{Frank}{Moss} \and \author{P.V.E.}{McClintock}}{1987}
            {Noise and nonlinear dynamical systems}
            {Cambridge University Press}{Cambridge}

\refarticle{refM104}
{\author{Miltiades}{Georgiou} \and \author{Thomas}{Erneux}}{1992}
          {Pulsating laser oscillations depend on 
        extremely-small-amplitude noise}
           {Phys. Rev. A}{45}{6636--6642}

\refbook{refM105}
{\author{A}{Atchison}}{19**}
            {The lognormal distribution}
            {Cambridge University Press}{Cambridge}

\refarticletobe{refM106}
{\author{Michael C.}{Mackey} \and \author{Irina G.}{Nechaeva}}{1993}
            {Noise and stability in differential delay equations}
             {Journal of Dynamics and Differential Equations}

\refarticle{refM107}
{\author{Christopher W.}{Meyer}, \author{Guenter}{Ahlers}
\and \author{David S.}{Cannell}}{1991}
            {Stochastic influences on pattern formation in
Rayleigh-B\'enard convection: Ramping experiments}
             {Phys. Rev. A}{44}{2514--2537}

\refbook{refM108}
{\author{Kenneth S.}{Miller}}{1974}
            {Complex Stochastic Processes}
            {Addison-Wesley}{Reading}
\refbook{refM208}
{\author{Kenneth S.}{Miller}}{1964}
            {Multidimensional Gaussian Distributions}
            {Wiley}{New York}

\refarticle{refM109}
{\author{R.}{Mannella}
\and \author{P.V.E.}{McClintock}}{1990}
           {Noise in nonlinear dynamical systems}
           {Contemporary Physics}{31}{179--194}

\refarticle{refM110}
{\author{P.C.}{Matthews}, \author{M.R.E.}{Proctor},
\author{A.M.}{Rucklidge} \and \author{N.O.}{Weiss}}{1993}
            {Pulsating waves in nonlinear magnetoconvection}
            {Phys. Lett. A}{183}{69--75}
 
\refarticle{refM111}
{\author{Frank}{Moss}, \author{D.K.}{Kondepudi}
 \and \author{P.V.E.}{McClintock}}{1985}
            {Branch selectivity at the bifurcation of a bistable
system with external noise}
            {Phys. Lett. A}{112}{293--296}
 
\refarticle{refM112}
{\author{Brian}{Morris} \and \author{Frank}{Moss}}{1986}
            {Postponed bifurcations of a quadratic map with a swept parameter}
            {Phys. Lett. A}{118}{117--120}
 
\refarticle{refM113}
{\author{C.}{Meunier} \and \author{A.D.}{Verga}}{1988}
            {Noise and Bifurcations}
            {J. Stat. Phys.}{50}{345--375}
 
\refarticle{refM114}
{\author{Michael C.}{Mackey}, \author{Andr\'e}{Longtin}
\and \author{Andrzej}{Lasota}}{1990}
            {Noise-Induced Global Asymptotic Stability}
             {J. Stat. Phys.}{60}{735--751}

\refarticle{refN2}
{\author{R.S.}{Mckay} \and \author{C.}{Tresser}}{1987}
            {Some flesh on the skeleton: the bifurcation structure
                of bimodal maps}
            {Physica}{27D}{412--422}

\refarticle{refN3}
{\author{M.}{Nagata}, \author{M.R.E.}{Proctor} \and
 \author{N.O.}{Weiss}}{1990}
            {Transitions to asymmetry in magnetoconvection}
            {Geophys. Astrophys. Fluid Dynamics}{51}{211--241}
 
\refarticle{refN2}
{\author{A.C.}{Newell} \and \author{J.A.}{Whitehead}}{1969}
            {Finite bandwidth, finite amplitude convection}
            {J.~Fluid Mech.}{38}{279--303}
 
\refarticle{refN4}
{\author{J.}{Niederl\"ander}, \author{M.}{L\"ucke} \and
 \author{M.}{Kamps}}{1991}
            {Weakly nonlinear convection: Galerkin model, numerical simulation,
             and amplitude equation}
            {Z.~Phys.~B -- Condensed Matter}{82}{135--141}

\refarticle{refN100}
{\author{A.I.}{Neishtadt}}{1987}
           {Persistence of stability loss for dynamical bifurcations}
           {Differential Equations}{23}{1385--1391}

\refarticle{refN101}
{\author{Bruce}{McNamara} and {Kurt}{Wiesenfeld}}{1989}
           {Theory of stochastic resonance}
           {Phys. Rev. A}{39}{4854--4869}

\refarticle{refN102}
{\author{Alan C.}{Newell}, \author{Thierry}{Passot}
 \and \author{Joceline}{Lega}}{1993}
            {Order parameter equations for patterns}
            {Ann. Rev. Fluid Mech.}{25}{399--453}

\refarticle{refN103}
{\author{Ali H.}{Nayfeh} \and \author{Nestor E.}{Sanchez}}{1989}
            {Bifurcations in a forced softening Duffing oscillator}
            {Int. J. Non-Linear Mechanics}{24}{483--497}
 
\refarticle{refN104}
{\author{S.}{Novak} \and \author{R. G.}{Frehlich}}{1982}
            {Transition to chaos in the Duffing oscillator}
            {Phys. Rev. A}{26}{3660--3663}
 
\refbook{refO100}
{\author{B.}{\O ksendal}}{1989}
    {Stochastic Differential Equations}
    {Springer}{Berlin}

\refbook{refP6}
{\author{T.S.}{Parker} \and \author{L.O.}{Chua}}{1989}
            {Practical Numerical Algorithms for Chaotic Systems}
            {Springer}{New York}
 
\refbook{refP2}
{\author{W.H.}{Press}, \author{B.P.}{Flannery},
 \author{S.A.}{Teukolsky} \and \author{W.T.}{Vetterling}}{1986}
            {Numerical Recipes -- the Art of Scientific Computing}
            {Cambridge University Press}{Cambridge}
 
\refarticle{refP4}
{\author{M.R.E.}{Proctor} \and \author{C.A.}{Jones}}{1988}
            {The interaction of two spatially resonant patterns in thermal
             convection. Part~1. Exact $1:2$~resonance}
            {J.~Fluid Mech.}{188}{301--335}
 
\refarticle{refP1}
{\author{M.R.E.}{Proctor} \and \author{N.O.}{Weiss}}{1982}
            {Magnetoconvection}
            {Rep. Prog. Phys.}{45}{1317--1379}
 
\refarticle{refP8}
{\author{M.R.E.}{Proctor} \and \author{N.O.}{Weiss}}{1990}
            {Normal forms and chaos in thermosolutal convection}
            {Nonlinearity}{3}{619--637}

\refartbook{refP100}
{\author{M.R.E.}{Proctor} \and \author{D.W.}{Hughes}}{1990}
        {Chaos and the effect of noise for the double Hopf
bifurcation with 2:1 resonance}
        {Nonlinear Evolution of Spatio-Temporal Structures in
Dissipative Continuous Systems}
        {F H Busse and L Kramer}
        {Plenum}{New York}{}
\refarticle{refP101}
{\author{F.}{de Pasquale}, \author{J.M.}{Sancho}, \author{M.}{San Miguel}
 \and \author{P.}{Tartaglia}}{1986}
            {Decay of an unstable state in the presence of
multiplicative noise}
            {Phys. Rev. A}{33}{4360--4366}

\refarticle{refP102}
{\author{P.}{Piera\'nski}
 \and \author{J.}{Ma\l ecki}}{1987}
            {Noise-sensitive Hysteresis Loops around Period-Doubling
Bifurcation Points}
            {Il Nuovo Cimento}{9}{757--779}
 
\refbook{refP103}
{\author{Guiseppe}{Da Prato} \and \author{Jerzy}{Zabczyk}}{1992}
    {Stochastic Equations in Infinite Dimensions}
    {Cambridge University Press}{Cambridge}     

\refbook{refP104}
{\author{Phoolan}{Prasad} \and \author{Renuka}{Ravindran}}{1985}
    {Partial Differential Equations}
    {Wiley Eastern}{New Delhi}  

\refarticle{refP105}
{\author{Nathan}{Platt}, \author{Stephen M.}{Hammel}
\and \author{James F.}{Heagy}}{1994}
            {Effect of Additive Noise on On-Off Intermittency}
            {Phys. Rev. Lett.}{52}{3498--3501}

\refbook{refP106}
{\author{Philip}{Protter}}{1990}
    {Stochastic Integration and Differential Equations}
    {Springer}{Berlin}

\refarticle{refR5}
{Lord~Rayleigh}{1916}
            {On convection currents in a horizontal layer of fluid, when the
             higher temperature is on the under side}
            {Phil.~Mag.}{32}{529--546}
 
\refarticle{refR13}
{\author{A.J.}{Rodr\'\i guez-Luis}, \author{E.}{Freire}
 \and \author{E.}{Ponce}}{1991}
            {On a codimension 3 bifurcation arising in an autonomous electronic
             circuit}
            {International Series of Numerical Mathematics}{97}{301--306}
 
\refarticle{refR8}
{\author{H.T.}{Rossby}}{1969}
            {A study of B\'enard convection with and without rotation}
            {J.~Fluid Mech.}{36}{309--335}
 
\refarticle{refR2}
{\author{A.}{Rucklidge} \and \author{S.}{Zaleski}}{1988}
            {A microcanonical model for interface formation}
            {J.~Stat. Phys.}{51}{299--307}
 
\refarticle{refR9}
{\author{A.M.}{Rucklidge}}{1992}
            {Chaos in models of double convection}
            {J.~Fluid Mech.}{237}{209--229}



\refbook{refR101}
{\author{Daniel}{Revuz} \and \author{Marc}{Yor}}{1991}
            {Continuous Martingales and Brownian Motion}
            {Springer}{Berlin}

\refbook{refR100}
{\author{H.}{Risken}}{1984}
            {The Fokker-Planck Equation}
            {Springer}{Berlin}
\refarticle{refR102}
{\author{L.A.}{Rubenfeld}}{1979}
            {A model bifurcation problem exhibiting the effects of
slow passage through critical}
            {SIAM J. Appl. Math.}{37}{302--306}
 
\refphd{refR103}
{\author{A.M.}{Rucklidge}}{1991}
            {Chaos in models of double convection}
            {University of Cambridge}
  
\refarticle{refR105}
{\author{A.M.}{Rucklidge} \and \author{P.C.}{Matthews}}{1995}
            {Analysis of the shearing instability
 in nonlinear magnetoconvection}
            {Nonlinearity}{9}{311--352}

\refarticle{refS31}
{\author{T.}{Sauer} \and \author{J.A.}{Yorke}}{1991}
            {Rigorous verification of trajectories for the computer simulation
             of dynamical systems}
            {Nonlinearity}{4}{961--979}
 
 

\refarticle{refS13}
{\author{L.P.}{Shil'nikov}}{1965}
            {A case of the existence of a countable number of periodic motions}
            {Soviet Maths. Dokl.}{6}{163--167}

\refarticle{refS25}
{\author{T.}{Shimizu} \and \author{N.}{Morioka}}{1978}
            {Chaos and limit cycles in the Lorenz model}
            {Phys. Lett.}{66A}{182--184}
 
\refarticle{refS18}
{\author{T.}{Shimizu} \and \author{N.}{Morioka}}{1980}
            {On the bifurcation of a symmetric limit cycle to an
             asymmetric one in a simple model}
            {Phys. Lett.}{76A}{201--204}

\refarticle{refS10}
{\author{T.G.L.}{Shirtcliffe}}{1969}
            {An experimental investigation of thermosolutal convection at
             marginal stability}
            {J.~Fluid Mech.}{35}{677--688}
 
\refarticle{refS19}
{\author{L.}{Sirovich} \and \author{A.E.}{Deane}}{1991}
            {A computational study of Rayleigh--B\'enard convection. Part~2.
             Dimension considerations}
            {J.~Fluid Mech.}{222}{251--265}
 
\refbook{refS1}
{\author{C.}{Sparrow}}{1982}
            {The Lorenz Equations: Bifurcations, Chaos, and Strange Attractors}
            {Springer}{New York}
 
\refarticle{refS9}
{\author{E.A.}{Spiegel}}{1987}
            {Chaos: a mixed metaphor for turbulence}
            {Proc. R.~Soc. Lond.}{413A}{87--95}
 
\refarticle{refS20}
{\author{J.W.}{Swift} \and \author{K.}{Wiesenfeld}}{1984}
            {Suppression of period doubling in symmetric systems}
            {Phys. Rev. Lett.}{52}{705--708}
 
\refphd{refS30}
{\author{J.}{Swinton}}{1991}
            {Stability of homoclinic waves in lasers and fluids}
            {Imperial College, London}
 
\refarticle{refS6}
{\author{P.}{Sz\'epfalusy} \and \author{T.}{T\'el}}{1985}
            {Properties of maps related to flows around a saddle point}
            {Physica}{16D}{252--264}
\refarticle{refS100}
{\author{Emily}{Stone} \and \author{Philip}{Holmes}}{1990}
            {Random perturbations of heteroclinic attractors}
            {SIAM J. Appl. Math.}{50}{726--743}

\refbook{refS101}
{\author{R}{Stratonovich}}{1963}
            {Topics in the theory of random noise}
            {Gordon and Breach}{New York}

\refarticle{refS102}
{\author{N.G.}{Stocks}, \author{R.}{Mannella} \and \author{P.V.E.}{McClintock}}{1989}
            {Influence of random fluctuations on delayed bifurcations:
        The case of additive white noise}
            {Phys. Rev. A}{40}{5361--5369}

\refarticle{refS103}
{\author{N.G.}{Stocks}, \author{R.}{Mannella} \and \author{P.V.E.}{McClintock}}{1990}
            {Influence of random fluctuations on delayed bifurcations.
        II. The cases of white and coloured additive 
        and multiplicative noise}
            {Phys. Rev. A}{42}{3356--3362}

\refartbook{refS104}
{\author{R.L.}{Stratonovich}}{1989}
            {Some Markov methods in the theory of stochastic processes
in nonlinear dynamical systems}
            {Noise in nonlinear dynamical systems, Vol. I}
            {\author{Frank}{Moss} \and \author{P.V.E.}{McClintock}}
            {Cambridge University Press}{Cambridge}{16--71}
 
\refartbook{refS105}
{\author{J.M.}{Sancho} \and \author{M.}{San Miguel}}{1989}
            {Langevin equations with coloured noise}
            {Noise in nonlinear dynamical systems, Vol. I}
            {\author{Frank}{Moss} \and \author{P.V.E.}{McClintock}}
            {Cambridge University Press}{Cambridge}{72--109}
 
\refarticle{refS106}
{\author{N.G.}{Stocks}, \author{N.D.}{Stein} \and \author{P.V.E.}{McClintock}}{1993}
            {Stochastic resonance in monostable systems}
            {J. Phys. A}{26}{L385--L390}

\refarticle{refS107}
{\author{David}{Sigeti} \and \author{Werner}{Horsthemke}}{1989}
            {Pseudo-regular oscillations induced by external noise}
            {J. Stat. Phys}{50}{1217--1222}

\refarticle{refS108}
{\author{Steven H.}{Strogatz}, \author{Renato E.}{Mirollo}
\and \author{Paul C.}{Matthews}}{1992}
            {Coupled Nonlinear Oscillators below the Synchronization
Threshold: Relaxation by Generalized Landau Damping}
            {Phys. Rev. Lett.}{68}{2730--2733}

\refarticle{refS109}
{\author{J.W.}{Swift}, \author{P.C.}{Hohenberg}
 \and \author{Guenter}{Ahlers}}{1991}
            {Stochastic Landau equation with time-dependent drift}
            {Phys. Rev. A}{43}{6572--6580}
 
\refarticle{refS110}
{\author{John}{Smythe}, \author{Frank}{Moss} \and
\author{P.V.E.}{McClintock}}{1983}
{Observation of a Noise-induced Phase Transition with an Analog Simulator}
            {Phys. Rev. Lett.}{51}{1062--1065}

\refarticle{refS111}
{\author{V.E.}{Shapiro}}{1993}
            {Systems near a critical point under multiplicative noise 
and the concept of effective potential}
            {Phys. Rev. E}{48}{109--120}

\refarticle{refS112}
{\author{V.E.}{Shishkova}}{1973}
            {Examination of a system of differential equations with a
small parameter in the highest derivatives}
            {Soviet Math. Dokl.}{14}{183--187}

\refbook{refS113}
{\author{Joel}{Smoller}}{1983}
            {Shock Waves and Reaction-Diffusion Equations}
            {Springer}{New York}

\refarticle{refS114}
{\author{D.J.}{Scalapino}, \author{M.}{Sears} \and
\author{R.A.}{Ferrell}}{1972}
        {Statistical Mechanics of One-Dimensional Ginzburg-Landau Fields}
         {Phys. Rev. B}{6}{3409--3416}

\refarticle{refS115}
{\author{Masuo}{Suzuki}}{1978}
        {Theory of instability, Brownian motion and formation of
macroscopic order}
         {Phys. Lett. A}{67}{339--341}
 
\refarticle{refT4}
{\author{F.}{Takens}}{1974}
            {Forced oscillations and bifurcations}
            {Comm. Math. Inst., Rijksuniversiteit Utrecht}{3}{1--59}
 
\refarticle{refV1}
{\author{G.}{Veronis}}{1965}
            {On finite amplitude instability in thermohaline convection}
            {J.~Marine Res.}{23}{1--17}
 
\refarticle{refV4}
{\author{G.}{Veronis}}{1966}
            {Motions at subcritical values of the Rayleigh number in a rotating
             fluid}
            {J.~Fluid Mech.}{24}{545--554}
 
\refarticle{refV7}
{\author{G.}{Veronis}}{1966}
            {Large-amplitude B\'enard convection}
            {J.~Fluid Mech.}{26}{49--68}
 
\refarticle{refV4a}
{\author{G.}{Veronis}}{1966a}
            {Motions at subcritical values of the Rayleigh number in a rotating
             fluid}
            {J.~Fluid Mech.}{24}{545--554}
 
\refarticle{refV7b}
{\author{G.}{Veronis}}{1966b}
            {Large-amplitude B\'enard convection}
            {J.~Fluid Mech.}{26}{49--68}
 
\refarticle{refV5}
{\author{G.}{Veronis}}{1968}
            {Large-amplitude B\'enard convection in a rotating fluid}
            {J.~Fluid Mech.}{31}{113--139}

\refarticle{refW12}
{\author{R.W.}{Walden}, \author{P.}{Kolodner},
 \author{A.}{Passner} \and \author{C.M.}{Surko}}{1985}
            {Travelling waves and chaos in convection in binary fluid mixtures}
            {Phys. Rev. Lett.}{55}{496--499}
 
\refarticle{refW16}
{\author{N.O.}{Weiss}}{1966}
            {The expulsion of magnetic flux by eddies}
            {Proc. Roy. Soc.}{293A}{310--328}
 
\refarticle{refW3}
{\author{N.O.}{Weiss}}{1981}
            {Convection in an imposed magnetic field. Part 1. The development
             of nonlinear convection}
            {J.~Fluid Mech.}{108}{247--272}
 
\refarticle{refW4}
{\author{N.O.}{Weiss}}{1981}
            {Convection in an imposed magnetic field. Part~2. The dynamical
             regime}
            {J.~Fluid Mech.}{108}{273--289}
 
\refarticle{refW3a}
{\author{N.O.}{Weiss}}{1981a}
            {Convection in an imposed magnetic field. Part~1. The development
             of nonlinear convection}
            {J.~Fluid Mech.}{108}{247--272}
 
\refarticle{refW4b}
{\author{N.O.}{Weiss}}{1981b}
            {Convection in an imposed magnetic field. Part~2. The dynamical
             regime}
            {J.~Fluid Mech.}{108}{273--289}
 
\refarticle{refW11}
{\author{N.O.}{Weiss}, \author{D.P.}{Brownjohn},
 \author{N.E.}{Hurlburt} \and \author{M.R.E.}{Proctor}}{1990}
            {Oscillatory convection in sunspot umbrae}
            {Mon.~Not. R.~Astr. Soc.}{245}{434--452}

\refbook{refW5}
{\author{S.}{Wiggins}}{1988}
    {Global Bifurcations and Chaos: Analytical Methods}
    {Springer}{New York}
 
\refarticle{refV100}
{\author{S.Ya.}{Vyshkind} \and \author{M.I.}{Rabinovich}}{1976}
            {The phase stochastization mechanism and the structure of
wave turbulence in dissipative media}
            {Sov. Phys. JETP}{44}{292--299}
\refarticle{refW100}
{\author{J.-M.}{Wersinger}, \author{J.M.}{Finn}
  \and \author{E.}{Ott}}{1980}
            {Bifurcation and ``strange'' behaviour in instability
saturation by nonlinear three-wave mode coupling}
            {Phys. Fluids}{23}{1142--1154}

\refnote{refNote1}
{The change of variables from the equations for
the amplitudes of the three wave modes used here differs  from
that of Hughes and Proctor; our $(x,y,z)$ corresponds to $(-x,w,y)$}

\refnote{refNote2}
{The distribution of a variable whose logarithm is normally distributed is 
sometimes called lognormal.
In an earlier publication[3], we were unaware of
this convention and used the term `lognormal` for the distribution of
the logarithm of $\ha \ln(n^2)$
 where $n$ is a Gaussian random variable with unit variance.}

\refnote{refNote3}{In an earlier publication~[12] we called this the
``log-normal'' distribution, being unaware that that term is
conventionally used for the distribution of a variable whose logarithm
is normally distributed}

\refnote{refNote4}{Hughes and Proctor (reference [5]) performed a
similar reduction. The variables $x,y,z$ used here are related to the
$A,B,C$ used by them via $x=A$, $y=\ha(B+C)$ and $z=\ha(B-C)$; $\delta
= \ha(1-\lambda)$. This choice of variables used in this work
simplifies analysis because of the symmetry (6)}

\refarticle{refT100}
{\author{M.C.}{Torrent} \and \author{M.}{San Miguel}}{1988}
           {Stochastic-dynamics characterization of delayed laser
threshold instability with swept control parameter}
           {Phys. Rev. A}{38}{245--251}

\refarticle{refT101}
{\author{M.C.}{Torrent}, \author{F.}{Sagu\'es} \and \author{M.}{San Miguel}}{1989}
           {Dynamics of sweeping through an instability: Passage-time
statistics for colored noise}
           {Phys. Rev. A}{40}{6662--6672}

\refarticle{refT102}
{\author{M.}{Tlidi}, \author{Paul}{Mandel}  \and
\author{R.}{Lefever}}{1994}
           {Localized Structures and Localized Patterns in Optical Bistability}
           {Phys. Rev. Lett}{73}{640--643}

\refbook{refW101}
{\author{David}{Williams}}{1991}
            {Probability with Martingales}
            {Cambridge University Press}{Cambridge}

\refarticle{refV100}
{\author{Jorge}{Vi\~nals}, \author{Hao-Wen}{Xi}
  \and \author{J.D.}{Gunton}}{1992}
            {Numerical study of the influence of forcing terms
and fluctuations near onset on the roll pattern in Rayleigh-B\'enard
convection in a simple fluid}
         {Phys. Rev. A}{46}{918-927}

\refarticle{refW102}
{\author{Jean-Marie}{Wersinger}, \author{John M.}{Finn}
  \and \author{Edward}{Ott}}{1980}
            {Bifurcations and strange behaviour in instability
saturation by nonlinear mode coupling}
         {Phys. Rev. Lett.}{44}{453--457}

\refartbook{refW103}
{\author{J.B.}{Walsh}}{1986}
            {An introduction to stochastic partial differential equations}
            {Ecole d'\'et\'e de probabilit\'es de St-Flour XIV}
            {\author{P.L.}{Hennequin}}
            {Springer}{Berlin}{pp266--439}

\refphd{refW104}
{\author{R.A.}{Weston}}{1989}
            {Lattice Field Theory and Statistical-Mechanical Models}
            {University of Cambridge}
 
\refarticle{refW105}
{\author{K.}{Wiesenfeld} \and \author{Bruce}{McNamara}}{1985}
            {Period-Doubling Systems as Small-Signal Amplifiers}
            {Phys. Rev. Lett.}{55}{13--15}

\refartbook{refW106}
{\author{K.}{Wiesenfeld}}{1989}
            {Period doubling bifurcations: what good are they?}
            {Noise in nonlinear dynamical systems, Vol. II}
            {\author{Frank}{Moss} \and \author{P.V.E.}{McClintock}}
            {Cambridge University Press}{Cambridge}{pp145--178}

\refarticle{refW107}
{\author{K.}{Wiesenfeld}}{1985}
            {Virtual Hopf phenomenon: A new precursor of
period-doubling bifurcations}
            {Phys. Rev. A}{32}{1744--1752}

\refarticle{refW108}
{\author{K.}{Wiesenfeld} \and \author{N.F.}{Pederson}}{1987}
            {Amplitude calculation near a period-doubling bifurcation:
        An example}
            {Phys. Rev. A}{36}{1440--1444}
\refarticle{refW109}
{\author{C.}{DeWitt-Morette} \and \author{K.D.}{Elworthy (eds)}}{1987}
            {New stochastic methods in physics}
            {Phys. Rep.}{77}{121--382}

\refbook{refW110}
{\author{L.C.G.}{Rogers} \and \author{David}{Williams}}{1987}
            {Diffusions, Markov Processes and Martingales.
        Vol 2: \ito\ calculus}
            {Wiley}{Chichester}

\refarticle{refY100}
{\author{Marc}{Yor}}{1989}
            {On stochastic areas and averages of planar Brownian motion}
            {J. Phys. A}{22}{3049--3057}

\refarticle{refZ100}
{\author{H.}{Zeghlache}, \author{Paul}{Mandel} \and \author{C.}{van den Broeck}
}{1989}
           {Influence of noise on delayed bifurcations}
           {Phys. Rev. A}{40}{286--294}

\refartbook{refZ101}
{\author{Paul}{Mandel}, \author{H.}{Zeghlache} \and
\author{T.}{Erneux}}{1989}
           {Time-dependent phase transitions}
        {Far from Equilibrium phase Transitions}
{\author{Luis}{Garrido}}{Springer}{Berlin}{pp218--236}

\endgroup
\inrefsfalse

\spaceandahalf

\windowcover

\def\theauthor{G.D.~Lythe}

\title{A noise-controlled dynamic bifurcation}

\damtpaddress
\signature

\abstract{
We consider a slow passage through a point of loss of stability.  If the passage is
sufficiently slow, the dynamics are controlled by additive random
disturbances, even if they are extremely small.
  We derive expressions for the `exit value' distribution when
 the parameter is explicitly a function of time and the dynamics are
 controlled by additive Gaussian noise.
We derive a new expression for the small correction introduced if the noise is
coloured (exponentially correlated).
There is good agreement with results obtained from simulation of sample
paths of the appropriate stochastic differential equations.
Multiplicative noise does not produce noise-controlled 
dynamics in this fashion.}

\publishedin{Nonlinear Science B {\bf 7} 147--158 (1994)}

\begingroup
\ShowTP
\endgroup

\pageno=1

\beginsect Introduction

Bifurcations are normally observed by slowly changing a parameter
(`sweeping') and waiting for a
qualitative change in behaviour. The standard theory of bifurcations, however,
does not have explicit time-dependence of
parameters\addref{refL102}{L102}.  Here we study
the effect of noise on the loss of stability in the simple equation
$$\dot y = gy.
\eqno\myeqno\myeqlabel{doty}$$
This is a  linearised dynamic bifurcation: a bifurcation in which
 the parameter $g$ is explicitly a function of time.

Normally it is said that loss of stability in \doty\
 is signalled by exponential growth of $|y|$ for $g>0$.
However if we continuously increase $g$ from $g<0$ and wait for  $|y|$
to exceed some
fixed value, we find that this occurs well after $g=0$\addref{refN100}{N100}.
 In this paper we set
$$g=\mu t \eqno\myeqno\myeqlabel{gemt}$$
 and let $y$ take (fixed) initial value $y_0$ at
some $g=g_0<0$.   Then the solution of \doty ,
$$y=y_0e^{{1 \over 2\mu} (g^2-g_0^2)},
\eqno\myeqno\myeqlabel{solndoty}$$
reveals that $|y|<|y_0|$  until $g=-g_0$,
regardless of the magnitude of $g_0$ and how small the sweep rate $\mu$ is.
However if $\mu$ is sufficiently small the dynamics enter a
noise-controlled \regime, in which
 the value of $g$ when $|y|$ reaches $|y_0|$
is a random variable controlled by the size of additive random
 disturbances\addthreerefs{refB100}{B100}{refB101}{B101}{refT100}{T100}.
This random variable we call the exit value; it is
more akin to a mean first passage time than
to the usual definition of a bifurcation point\addref{refS102}{S102}.

This work is chiefly concerned with the probability 
distribution of the exit value 
in the ``noise-controlled \regime''\addref{refL100}{L100}
when \doty\ is perturbed by additive white and coloured noise.
Because \doty\ is a non-autonomous linear equation,
it is possible to calculate an exact solution for the \pd\ of $y$ as a
function of time and thus to calculate
 the probability that $|y|>|y_0|$, which we will denote by
${\rm Pr}(|y| >|y_0|)$.  The \pd\ of the exit value 
 is the derivative with respect to $g$ of ${\rm Pr}(|y| >|y_0|)$.
Additive noise reduces the average exit value by
defining a level below which $\mean{y^2}$ does not fall. 

The noise-controlled \regime\ is in force if 
$$\mle < \ha g_0^2 +\Or{\mu}
\eqno\myeqno\myeqlabel{nccond}$$
 where $\ep$ is the (r.m.s.) noise level.  This is the most
likely \regime\ for small $\mu$.
Concentrating on this \regime\ makes possible a simpler
analytical treatment than previously
 reported\addtworefs{refZ100}{Z100}{refS102}{S102},
including explicit derivation of the \pd\ of exit values.
This is presented in section 2.
We have chosen $|y|=|y_0|$ as the fixed exit level; this makes
transparent the difference between the noise-controlled
\regime\ and  the deterministic limit, but we emphasise that, 
as long as the noise-controlled \regime\ remains in force,
 the dynamics are independent of the initial
conditions.

Our analytical results are in excellent agreement with results
obtained from simulation of sample paths.  These simulations we
performed using an algorithm with weak second order convergence
(the errors in the moments of the exit value distribution are
proportional to the second power of the
timestep\addref{refK100}{K100}).
Our results are also consistent with the reported results of
analogue and digital
simulations\addtworefs{refM102}{M102}{refS102}{S102}
performed in  different parameter ranges.

If the noise is coloured, the \pd\ of $y$ is narrowed and the mean
exit value is increased\addtworefs{refZ100}{Z100}{refT100}{T100}.
The correction is, however, small except for very large values of the
correlation time $\tau$\addref{refS103}{S103}. In section 3 we
derive a new formula for this correction which we successfully compare
with numerical results in the (physically realistic) \regime\
$\mu\tau^2<\Or{1}$.
In section 4 we show that multiplicative noise has a qualitatively
different (and in general much less dramatic) effect on the dynamics.
In section 5 we discuss the situation where $y$ has spatial degrees of
freedom and the passage through the point of stability loss 
causes a growth of spatial order.

Noise-controlled dynamics are useful in understanding the
behaviour of systems of ordinary differential equations used as
low-order models in fluid
mechanics\addtworefs{refH103}{H103}{refL101}{L101}
 and laser physics\addtworefs{refM100}{M100}{refM104}{M104}.
In these cases the time-dependence of 
the parameter corresponding to $g$ is in general more complicated;
our system serves as a linearised model which, being a good
approximation near $g=0$, captures the essential dynamics.

\beginsect Additive white noise

To study the effect of additive white noise 
we replace the ordinary differential equation
 \doty\ by the stochastic differential equation (SDE)
$$dy_t = gy_tdt + \epsilon dW_t
\eqno \myeqno \myeqlabel{sdewhite}$$
where $\ep$ is constant with $0\le \ep \ll 1$ and $W_t$ is the Wiener
process\addref{refG100}{G100}. The  solution of this SDE with $g=\mu t$ is
$$y_t = e^{\ha\mu(t^2-t_0^2)}\left(y_0 + \epsilon
\int_{t_0}^{t}e^{-\ha\mu(t^2-s^2)}dW_s\right)
\eqno \myeqno \myeqlabel{dbsdesoln}$$
where $t_0=g_0/\mu$.
For each $t$, $y$ is a Gaussian random variable with mean
equal to the solution \solndoty\ of the corresponding deterministic equation
\doty\ and variance a function of time given by
$$\sigma^2 = \mean{y_t^2}-\mean{y_t}^2 = \ep^2e^{{1 \over \mu}g^2}
\int^t_{t_0}e^{-\mu s^2}ds.
\eqno \myeqno \myeqlabel{awvar}$$
If $-g_0 > \Or{\sqrt{\mu}}$ then, for $g>\Or{\sqrt{\mu}}$,
$$\sigma^2 = \ep^2e^{{1 \over \mu}g^2}\sqrt{{\pi \over \mu}}.
\eqno \myeqno \myeqlabel{awvarlt}$$

The noise-controlled \regime\ is in force 
when $\sigma^2 \gg  \mean{y_t}^2$ for
$g>\Or{\sqrt{\mu}}$, so that all memory of the initial
conditions $(g_0,y_0)$ is lost and
noise controls the exit value distribution.  For this we need
$$\mu|\ln \epsilon| <
{1 \over 2}g_0^2 +\mu \left ( {1 \over 4}\ln{\pi \over \mu}-\ln y_0
\right ).
\eqno \myeqno \myeqlabel{dbtrans}$$
Since the right hand side is $\Or{1}$ in general,
\dbtrans\ holds even for extremely small values of $\ep$
if $\mu$ is moderately small ($\mu < 0.1$)\addref{refL100}{L100}.

In what follows, we assume that the noise level, $\ep$,
 is $\ll |y_0|$ but large enough
that \dbtrans\ holds. For additive noise this means that, for
$g>\Or{\sqrt{\mu}}$, we can set $\mean{y_t}=0$.
The \pd\ of exit values is then given by
$$\eqalign{ {d \over dg}{\rm Pr}(|y| >|y_0|)
&= {d \over dg}\sqrt{2 \over \pi}{1 \over \sigma}\int_{y_0}^\infty
e^{{-x^2 \over 2\sigma^2}}dx\cr
&=\sqrt{2 \over \pi}{1 \over \mu\sigma^2}{d\sigma \over dt}
\left(-\int_{y_0}^\infty e^{{-x^2 \over 2\sigma^2}}dx+
\int_{y_0}^\infty {x^2 \over \sigma^2}e^{{-x^2 \over 2\sigma^2}}dx\right)\cr
&=\sqrt{2 \over \pi}{g \over \mu}{y_0 \over \sigma}e^{-y_0^2 \over 2\sigma^2}.}
\eqno \myeqno \myeqlabel{addg}$$
using integration by parts.

Let $\hat g$ be the value of $g$ such that $\sigma^2(\hat g) = y_0^2$
and let $v = -(g-\hat g)(\hat g/ \mu)$. Then
$\sigma \simeq e^v$ and the probability that
the exit value of $g$ lies between $g$ and $g+dg$
for additive white noise is $R_A(g)dg$ where,
to lowest order in $\mu$,
$$R_A(g) \simeq \sqrt{{2 \over \pi}}e^ve^{\ha e^{2v}}.
\eqno \myeqno \myeqlabel{pdexp}$$
This distribution is compared with numerical results in Figure 1;
it is the \pd\ of $\ha \ln n^2$
 where $n$ is a Gaussian random variable with unit variance.
The most probable value of $v$ is $0$, so
the most probable exit value is $\hat g$ where
$$\hat g = \sqrt{2\mu |\ln \epsilon| -{\mu \over 2}\ln{\pi \over\mu}+2\mu
\ln y_0}.\eqno \myeqno \myeqlabel{mpvg}$$
\figure 5.4truein  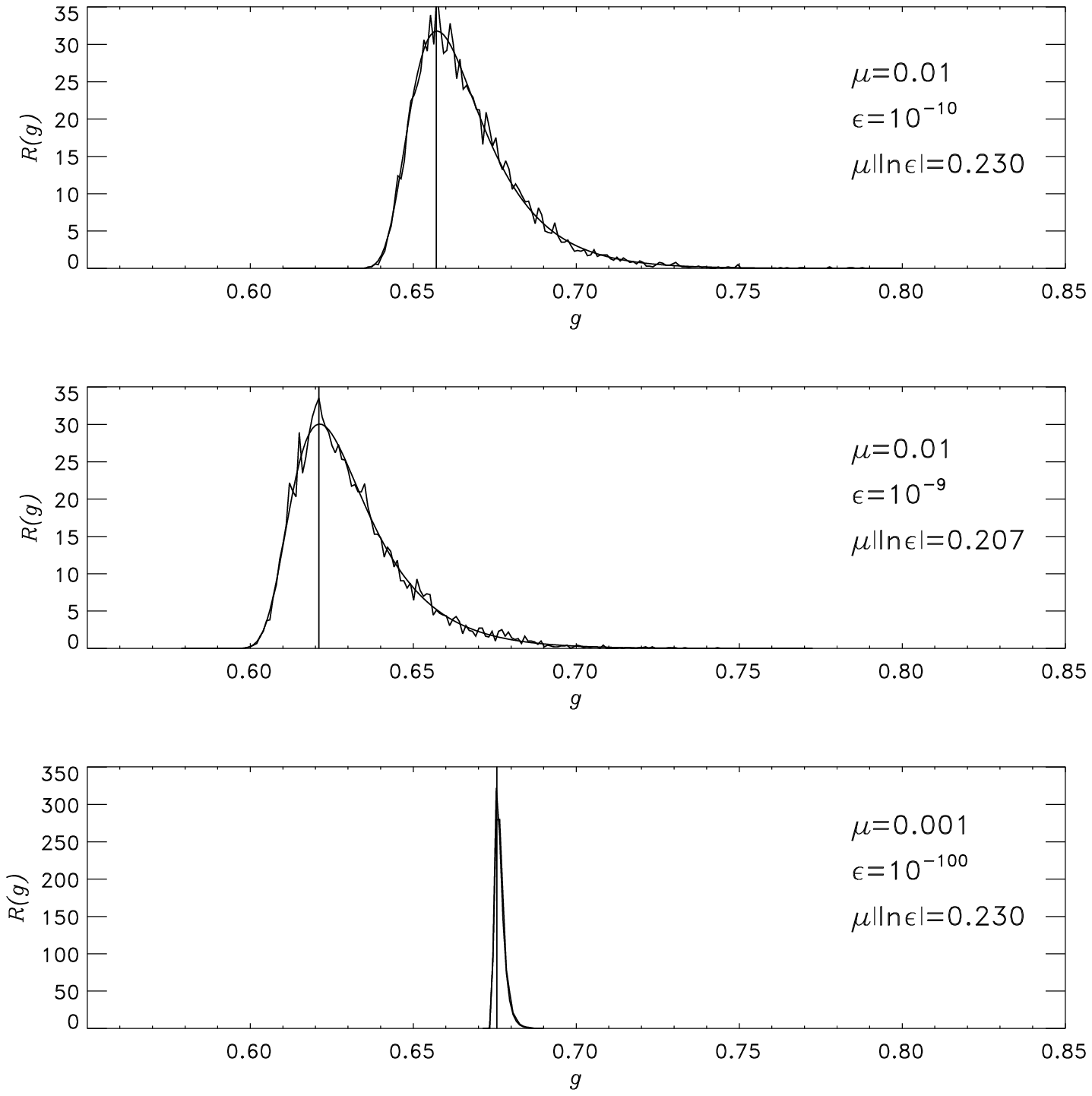
{\it Distribution of exit values in the noise-controlled \regime\ with  white noise.}
  Sample paths of \sdewhite\  were simulated numerically numerous times,
 with $y_0=1$ and $g_0 =-1$.
The next value of $g$ at which $y=1$ was recorded each time and the
resulting
distributions of exit values
 are plotted along with the calculated distribution \pdexp.
  The most probable value, given by \mpvg, is indicated by a vertical
line. (Without noise, the exit value is always $1$; if $\mle>0.5+
(\mu/ 4)\ln(\pi/\mu)$ the noise-controlled \regime\ is not in force
and the exit value distribution is narrow,
Gaussian and centred on $g=1$.)
\myfigurelabel{figaddwhite}

We can obtain exact expressions for the moments of the exit value
distribution in the noise-controlled \regime\
by noting that $v=\ha(v'+\ln 2)$ where $v'=\ln \ha n^2$.
The characteristic function, $\phi(s)$, of $v'$ is
known\addref{refR101}{R101}:
$$\phi(s)={1 \over \sqrt{\pi}}\int_{-\infty}^{\infty}e^{-e^u}e^{{1
\over 2}u}e^{isu}du = {1 \over \sqrt{\pi}}\Gamma({1 \over 2} + is).
\eqno \myeqno \myeqlabel{genfnv}$$
The mean exit value, $\mean{g_e}$ is 
$$\mean{g_e} = \hat g + {\mu\over 2\hat g}(\gamma +\ln 2)
\eqno \myeqno \myeqlabel{meanex}$$
where $\gamma=0.57721 \ldots$ (Euler's constant) and
and the variance of the distribution of exit values
is 
$$\mean{g_e^2}-\mean{g_e}^2 = {\pi^2\over8}{\mu^2\over\hat g^2}.
\eqno \myeqno \myeqlabel{varex}$$

It is instructive to consider how these results arise in terms of
trajectories (sample paths) of \sdewhite.
Firstly note that, once $g>\Or{\sqrt{\mu}}$, most trajectories have $|y|\gg\ep$,
and so are no longer affected by the noise.
  Additive noise acts near $g=0$, effectively
providing a random initial condition for the subsequent 
evolution\addtworefs{refP101}{P101}{refT100}{T100}.
The most probable exit value is determined by the time,
$\Or{\sqrt{\ln(y_0/\ep)/\mu}}$, taken for
the variance $\mean{y^2}$ to rise from a level proportional to $\ep^2$ at
$g=0$ to $y_0^2$ at $g=\hat g$.  
The \pd\ of exit values has width $\Or{\mu}$ because most members
of the ensemble of trajectories cross $|y|=|y_0|$ within a time $\Or{1}$
of each other, corresponding to an $\Or{\mu}$ spread in $g$.
Now consider the exponential rise of neighbouring trajectories;
 the difference in their exit values is proportional to the log
of the ratio of their starting points.  The Gaussian distribution
in $y$ near $g=0$ is thus transformed into a distribution of exit values
whose characteristic shape is that of $\ha\ln n^2$, where $n$ is a
Gaussian random variable.

\beginsect Additive coloured noise

Noise is often added to an ordinary differential equation
 as a way of approximating the influence
of neglected small effects while retaining simplicity of
description.
  Often ``coloured noise'', i.e. noise with non-zero correlation time, is more appropriate
than white noise.
  If $y$ is subject to coloured noise, we replace \sdewhite\ by
$$dy_t = \mu t y_tdt + \ep\eta_t dt.
\eqno \myeqno \myeqlabel{colsde}$$
For $\eta_t$ we use the most commonly studied coloured noise process\addref{refA100}{A100}, the \ou\
satisfying the SDE
$$d\eta_t = {\eta_t \over \tau}dt + dW_t
\eqno \myeqno \myeqlabel{etasde}$$
which has
$$\mean{\eta_t}=0 \qquad {\rm and}\qquad
\mean{\eta_s\eta_t} = {1 \over 2\tau}e^{-|s-t|/\tau}\quad \forall s,t.
\eqno \myeqno \myeqlabel{meaneta}$$
The solution of \colsde\ with $y=1$ at $t=t_0$ is
$$y_t =
e^{\ha\mu(t^2-t_0^2)}+\ep\int_{t_0}^{t}
e^{-\mu(t^2-s^2)}\eta(s)ds.
\eqno \myeqno \myeqlabel{colsoln}$$
We restrict ourselves to the noise-controlled \regime\ where, for $t>
\Or{{1 \over \sqrt{\mu}}}$,
$$\eqalign{\mean{y_t^2} &\simeq \ep^2e^{\mu t^2}{1 \over \tau}
\int_{t_0}^{t}\int_{t_0}^{s} e^{-{\mu\over 2}(s^2+u^2)-(s-u)/\tau}duds\cr
&= \ep^2e^{\mu t^2}{1 \over \tau}
\int_{t_0}^{t} e^{-{\mu\over 2}s^2-s/\tau+{1 \over 2\mu\tau^2}}
\sqrt{{2 \over \mu}}\int^{\alpha}_{\alpha_0}e^{-v^2}dvds}
\eqno \myeqno \myeqlabel{colmysc}$$
where $\alpha = \sqrt{{\mu/2}}(s- (\mu\tau)^{-1})$
and $\alpha_0 = \sqrt{{\mu/2}}(t_0- (\mu\tau)^{-1})$.
Most of the support of the integrand is near $s=0$.
We therefore use the following approximation for the error
function\addref{refA101}{A101},
$$\int_{\alpha_0}^{\alpha}e^{-v^2}dv
\simeq {e^{-\alpha^2} \over -\alpha +\sqrt{\alpha^2 + {4 \over \pi}}},
\eqno \myeqno \myeqlabel{apperrf}$$
and expand the denominator in \apperrf, about $s=0$, in a form
convenient for the integration over $s$:
$$\sqrt{{2\over \mu}}{1 \over \tau}{1 \over 
-\alpha +\sqrt{\alpha^2 + {4 \over \pi}}}
\simeq {2 \over 1+\kappa}e^{{\mu\tau  \over \kappa}s}
\eqno \myeqno \myeqlabel{mysexp}$$
where $\kappa = \sqrt{1+{8 \over \pi}\mu\tau^2}$.
Thus for $t-\tau$ and $|t_0|-\tau > \Or{{1/\sqrt{\mu}}}$, $\mean{y_t^2}$ is
approximately given by
$$\mean{y_t^2} \simeq \sqrt{{\pi\over \mu}}\ep^2e^{\mu t^2}
e^{\mu\tau^2 \over 4\kappa^2}{2 \over 1+\kappa}.
\eqno \myeqno \myeqlabel{colmys}$$
This is smaller than the white noise value \awvarlt\ by a factor which is a
function of $\mu \tau^2$.
The solution of \colsde\ also differs from that of \sdewhite\
in multi-time correlations $\mean{y_ty_s}$ but
this is not important for the purposes of calculating the
distribution of exit values of $g$. We
 make the approximation that noise with correlation time $\tau$
is equivalent to white noise at a slightly smaller level
$\ep'$ where
$$\ep' \simeq \ep \sqrt{2 \over 1+\kappa}e^{\mu\tau^2/ 8\kappa^2}.
\eqno \myeqno \myeqlabel{colres}$$
The mean exit value in the presence of exponentially correlated noise
with correlation time $\tau$ is therefore given by:
$$\mean{g_e} = \sqrt{2\mle -\mu f(\mu\tau^2)}+
{\mu \over 2\hat g}(\gamma+\ln 2) +\Or{\mu^2}
\eqno \myeqno \myeqlabel{colmean}$$
where
$$f(\mu\tau^2)=\ln({2 \over 1+\kappa})
-{\mu\tau^2\over 4\kappa^2}-{1 \over 2}\ln({\pi\over\mu}),
\eqno \myeqno \myeqlabel{fmts}$$
$\gamma=0.57721\ldots$ (Euler's constant) and $\hat g$ is given by
\mpvg.  This formula is compared with
numerical results in Figure 2. The 
most probable exit value and variance of the exit
value distribution can be calculated similarly.

\figure 5.6truein 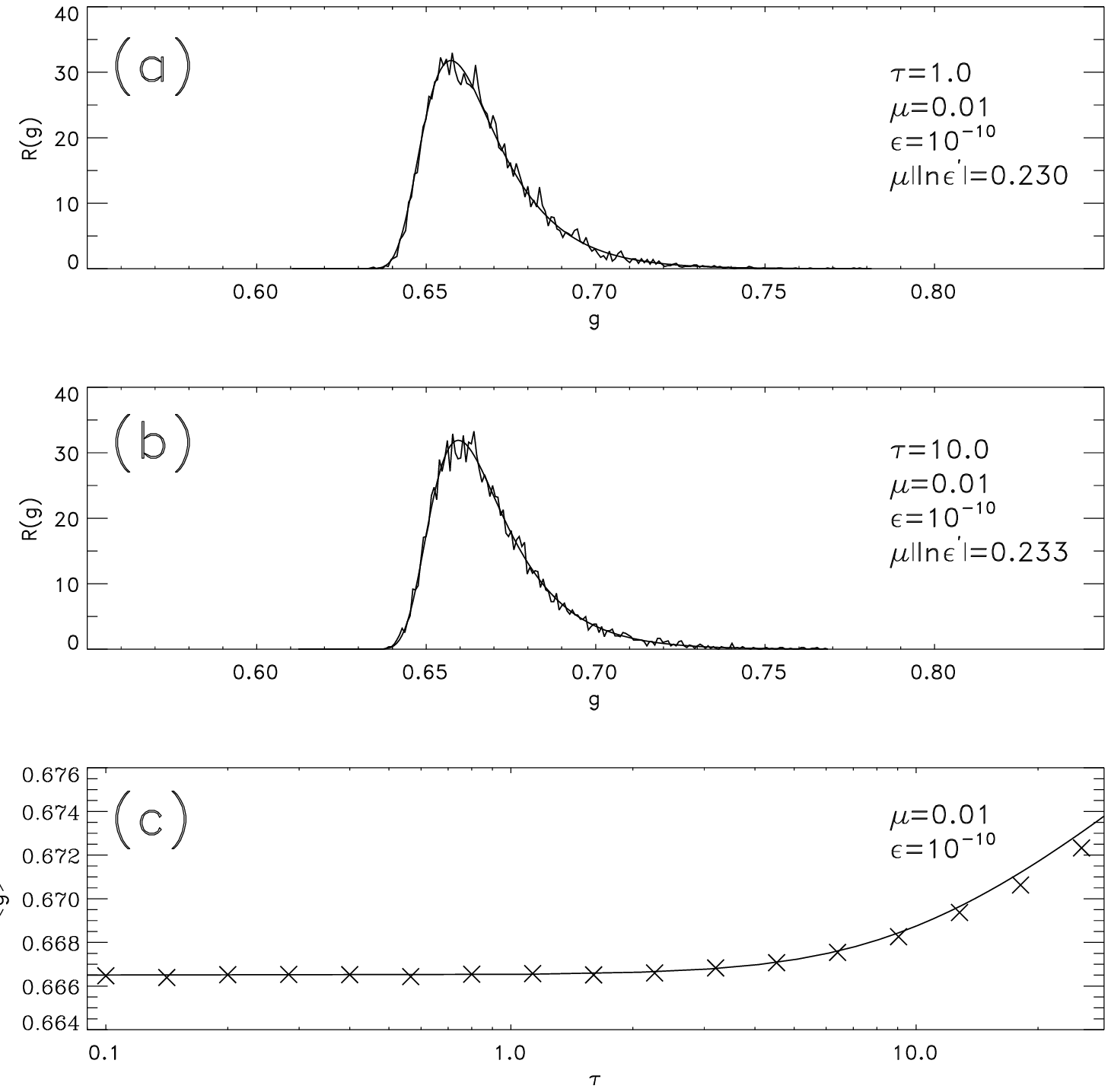
{\it Distribution of exit values in the noise-controlled \regime\ with coloured noise}.
In (a) and (b) the distribution obtained by simulation of sample paths
solution of \colsde\
is shown along with the calculated distribution \pdexp\
(smooth curve) calculated with $\ep$ replaced by $\ep'$, according to \colres,
to account for the effect of non-zero correlation time in the noise.
Graph (c) shows the calculated mean exit value as a function of
$\tau$ \colmean\ with $\mu=0.01$ and $\ep=10^{-10}$ (solid line)
and some numerical results (crosses).
\myfigurelabel{colfig}

\beginsect Multiplicative white noise

If $y$ is subject to noise which is proportional
to $y$ then we replace \doty\ by the SDE
$$dy = gy_tdt + \epsilon y_t dW_t.
\eqno \myeqno \myeqlabel{multsde}$$
(The most natural way for multiplicative noise
 to arise is if $g$ itself is subject to fluctuations.)
The solution of \multsde\ with $y=y_0$ at $t-t_0$
is \addref{refG100}{G100}
$$y=y_0e^{\ha\mu(t^2-t_0^2)-\ha\epsilon^2(t-t_0)}
e^{\epsilon (W_t-W_{t_0})}
\eqno \myeqno \myeqlabel{multsoln}$$
where $t_0=g_0/\mu$.
The median value of $y$ is $y_0\exp({1 \over
2\mu}(g^2-g_0^2-\ep^2(g-g_0)))$;
the factor $\exp(\ep(W_t-W_{t_0}))$ is always positive with mean
$\exp(\ha\ep^2(t-t_0))$.
As before, we first calculate
the probability that $y>y_0$ as a function of $g$.
$$Pr(|y|>|y_0|) = Pr(e^{\ep (W_t-W_{t_0})}>{1 \over m})={1 \over \sqrt{2\pi\sigma_m^2}}\int_{-\ln m}^{\infty}
e^{-{x^2 \over 2\sigma_m^2}}dx
\eqno \myeqno \myeqlabel{multgdistn}$$
where $m=\exp(\ha\mu(t^2-t_0^2)-\ha \ep^2(t-t_0))$
and $\sigma_m^2 = \ep^2(t-t_0)$.
When $\ep\ll1$ the possibility that one trajectory crosses $|y|=|y_0|$
more than once can be ignored; the exit value distribution is then
Gaussian, centred on $-g_0+\ep^2$, with
variance $(2\mu/ |g_0|)\ep^2$.

Using our exit value definition, we find that multiplicative noise
increases the median delay even beyond the deterministic value.
Other definitions, based for example on the trajectory of 
$\mean{y_t^2}$\addtworefs{refB101}{B101}{refZ100}{Z100},
may lead to the opposite conclusion because
$\mean{y_t}$ and $\sqrt{\mean{y^2_t}}$ are greater than the median
value of $y$.  Regardless of definition, the effect is small if $\ep$ is small.
For multiplicative noise, there is no analogue of the
noise-controlled \regime;
we need  $\Or{1}$ multiplicative noise to produce $\Or{1}$ effects.
 We have demonstrated this when the noise is
simply proportional to $y$; it will remain true if the noise is any
$\Or{1}$ function of $y$ because the noise also becomes tiny as $y$
does, and therefore does not dominate at any stage.
In the presence of both additive and multiplicative noise, an exact
solution can still be found for the \pd\ of $y$ as a function of time;
to derive the exit value statistics, it is useful to note that
additive noise acts primarily near $g=0$ and multiplicative noise 
later\addtworefs{refP101}{P101}{refS103}{S103}.

\beginsect Conclusion

When the sweep through the point of stability loss is slow, the dynamics
 are controlled by very small random additive
noise and memory of initial conditions is lost.
For small values of the sweep rate ($\mu < 0.1$), noise
 produces $\Or{1}$ effects even if its magnitude is extremely small.
  In this noise-controlled \regime\ the most probable value of
$g$ when $y$ reaches an $\Or{1}$ predetermined level
 is $\hat g = \sqrt{2\mle}+\Or{\mu}$ 
and the \pd\ of exit values has standard deviation proportional to $\mu$.
As long as the noise is additive and Gaussian, its correlation time
 affects the exit value distribution only slightly.
  Multiplicative noise does not
produce noise-controlled dynamics in this way. 

A physical picture of this noise-controlled system
 is the dynamics  of a strongly
damped object allowed to slide in a slowly-changing potential
and continuously subject to random influences.
The variable $y$ is the distance from the extremum of the potential.
For $t<0$, the extremum
 is the centre of a well; for $t>0$ the extremum is the crest of a
hill.  Once $t>0$, the 
 object lingers near the crest for a time controlled by the noise level before sliding
away.

\beginack
\noindent I am grateful for financial support from the Commonwealth
Scholarship Commission, from Trinity College and from the department of applied
mathematics and theoretical physics.

\smallskip

\showrefs

\Trailers

\end